 \documentclass[final,3p,times]{elsarticle}

\usepackage{graphicx, amsmath, amssymb, mathtools, dirtytalk, bm, float} 
\usepackage[symbol]{footmisc}
\usepackage{color}
\usepackage{hyperref}
\usepackage{algpseudocode}
\usepackage{algorithm}
\usepackage{comment}
\usepackage{tikz}
\usetikzlibrary{decorations.markings}

\newcommand{\RR}{{\mathbb R}}
\newcommand{\ZZ}{{\mathbb Z}}
\newcommand{\OO}{{\mathcal O}}

\def\Xint#1{\mathchoice
   {\XXint\displaystyle\textstyle{#1}}%
   {\XXint\textstyle\scriptstyle{#1}}%
   {\XXint\scriptstyle\scriptscriptstyle{#1}}%
   {\XXint\scriptscriptstyle\scriptscriptstyle{#1}}%
   \!\int}
\def\XXint#1#2#3{{\setbox0=\hbox{$#1{#2#3}{\int}$}
     \vcenter{\hbox{$#2#3$}}\kern-.5\wd0}}

\def\dashint{\Xint-}
\newcommand{\jtc}[1]{{\color{blue}#1}}

\journal{Journal of Computational Physics}

\begin{document}

\begin{frontmatter}



\title{A split-step Christov method for approximating rational PDE solutions}


\author[label1]{Justin T. Cole}
\author[label1]{Troy I. Johnson}

\affiliation[label1]{organization={University of Colorado, Colorado Springs},
            addressline={1420 Austin Bluffs Parkway}, 
            city={Colorado Springs},
            postcode={80918}, 
            state={CO},
            country={USA}}

\begin{abstract}

Rational solutions of partial differential equations (PDEs)   are notoriously difficult to approximate via spectral Fourier methods due to their algebraically slow decay rate.
In this work we discuss approximating rational PDE solutions 
in a basis of orthogonal functions known as the 
Fourier series, allowing for the computation of its spectrum via the fast Fourier transform. Spectral differentiation matrices are derived. Several explicit fourth-order split-step integrators are derived and their performance compared.
As an application, rogue wave solutions in a family of nonlinear Schr\"odinger equations are explored. 
Perturbing the constant background is found to generate rogue wave-like structures. 
The effects of higher-order dispersion and generalized nonlinearities are also examined.

\end{abstract}

\begin{graphicalabstract}
\end{graphicalabstract}

\begin{highlights}
\item Approximation of rational solutions to partial differential equations by 
a  family of orthogonal rational functions called the 
Christov functions. For certain rational functions, these approximations converge spectrally (exponentially) fast. Spectral differentiation matrices are derived.

\item Three different explicit fourth-order split-step methods are derived and their performance compared: Yoshida, partitioned Runge-Kutta, Runge-Kutta Nystr\"om. The latter two are found to be superior.

\item Rogue wave models in a family of nonlinear Schr\"odinger equations are studied, including the Peregrine soliton. Perturbation of the constant background generates rogue wave-type structures which are  rational in nature. 
\end{highlights}

\begin{keyword}
 Christov functions \sep split-step integrators \sep rational functions \sep rogue waves



\end{keyword}

\end{frontmatter}


\section{Introduction}
\label{intro_sec}



%

The industry standard for numerically approximating  wave-related (that is, periodic) phenomena is spectral Fourier methods. For relatively smooth solutions, Fourier spectral methods are  an attractive option due to their efficient computation  (via fast Fourier transform) and rapid convergence rates \cite{Fornberg1996,Canuto2006}.  Fourier approximations can also be useful when approximating exponentially decaying solutions on the real line. In the latter case, one typically  truncates the infinite line to a large but finite domain on which an exponentially localized solution can be treated as a periodic function \cite{Fornberg1978,Yang2010}. However, this approach requires the function reach its 
boundary conditions well within the computational window. 

On the other hand, localized rational functions are  not well-suited for approximation by Fourier methods due to their algebraically slow decay rate.  On most feasible computational domains, rational functions are nowhere near their boundary conditions at the endpoints. 
As a result, their periodic extensions are not sufficiently smooth  and so, in general, their Fourier coefficients decay algebraically (slow). 


In order to better approximate rational functions several ideas have been explored. One  is to use the Hermite functions \cite{Boyd1980,Guo2003} as a basis. This is a natural approach since the Hermite functions form an orthogonal basis on the real line. However,  there is no known  fast Hermite transform and the truncated series only converges spectrally fast if the function decays exponentially fast at $\pm \infty$ \cite{Canuto2006}. Another approach is that of coordinate mapping \cite{Grosch1977,Islas2017}. The idea here is to map the (infinite) real line to a finite interval by some coordinate transformation. One such mapping is the so-called algebraic type, e.g. $x = \frac{ y}{\sqrt{1 - y^2}}$ for $y \in (-1,1)$ \cite{Boyd2001}. A method that uses said mapping and approximates rational solutions 
is the Chebyshev pseudo-spectral method given in \cite{Schober2017}. 

In this work we propose 
a {\it rational} basis to approximate {\it rational} functions.
A family of orthogonal rational functions was considered in  \cite{Boyd1987} that consists of the Chebyshev polynomials evaluated on the algebraic mapping domain, 
i.e. $\psi_n(x) = T_n\left( \frac{x}{\sqrt{1 + x^2}} \right), n = 0,1,\dots$ for $x \in (-\infty ,\infty)$, where $T_n $ are the Chebyshev polynomials. This orthogonal basis yields exponential convergence for functions that are analytic on the real line, as well as  banded differentiation matrices. Another candidate for a rational basis is the so-called  
Christov (Ch) functions \cite{Malmquist1925,Takenaka1926,Christov1982}. It was  shown in \cite{Boyd1990} that the Ch functions  are actually equivalent to the Chebyshev polynomials through a linear combination. As such, they inherit the many desirable properties mentioned above. Moreover, this family of functions are mutually 
orthonormal and  the coefficients can be computed via   Fourier transform.

In this work we develop a split-step 
Ch approach for approximating rational partial differential equation (PDE) solutions; this is the primary contribution of this work. The accuracy and efficiency of several explicit and symplectic fourth-order time-integrators 
are compared. Namely, we examine the Yoshida \cite{Yang2010,Yoshida1990}, partitioned Runge-Kutta, and Runge-Kutta Nystr\"om integrators \cite{McLachlan2021,Blanes2000}.
For benchmark 
simulations considered in this work, we are 
consistently able to achieve seven to eight  decimal places of accuracy in only a few seconds on a standard personal laptop.

As an application, 
we approximate rogue wave solutions of the  nonlinear Schr\"odinger (NLS) equation, an extensively studied integrable PDE model that has been derived in the context of water waves \cite{Zhakharov1968}, fiber optics \cite{Agrawal2013}, and plasmas \cite{Zhakharov1972}, to name a few. Rogue wave-like structures are generated by perturbing the constant background \cite{Kharif2003,Chabchoub2011}. 
In one instance, exact 
rational solutions 
are localized in space and time are used to model rogue wave behavior \cite{Yang2012,Yang2021}, the most well-known being the Peregrine soliton \cite{Peregrine1983}. A second case involves 
modulational instability (MI) where perturbations with a small sideband wavenumber are linearly unstable \cite{Benjamin1967,Benjamin1967b} and generate spatially localized, temporally recurrent structures. \cite{CuevasMaraver2017,Klein2017,Ablowitz2021}. 

The MI route is worth pursuing, but it should be performed on the real line and not on a periodic domain, to truly capture it's nature in large bodies of water, for example.  Recently, several works have highlighted the unstable nature of the Peregrine soliton.  Conversely, Peregrine-like localized structures are observed in recurrent MI dynamics \cite{Grimshaw2013}. This work examines both models. 

Later, we examine rogue wave formation in several non-integrable variants of the NLS equation. These extensions include generalized nonlinearity and higher-order dispersion, physical effects typically neglected in the NLS model. These experiments show that the higher-order nonlinearity dominates the dynamics and third-order dispersion can affect the formation of genuine rogue waves.


The outline of the work is as follows. In Section 2 we introduce the 
Ch functions and some of their properties. We discuss an expansion in terms of these functions and the rapid decay of their coefficients. In Section 3, spectral Galerkin differentiation matrices are constructed to approximate derivatives. In Section 4 we develop a novel spectral Galerkin method for approximating 
PDEs. In particular, we establish a split-step approach to approximate 
the dynamics of the NLS equation. 
In Section 5 we explore instability in the Peregrine soliton and mechanisms for generating rogue waves on a constant background. In Section 6 we apply this scheme to non-integrable variants of the NLS equation which include generalized nonlinearity and higher-order dispersion. We conclude in Section 7. Several appendices containing technical notes are also included.

\section{The Christov functions}
\label{Ch_sec}

\begin{figure}
 \centering
 \includegraphics[scale=0.45]{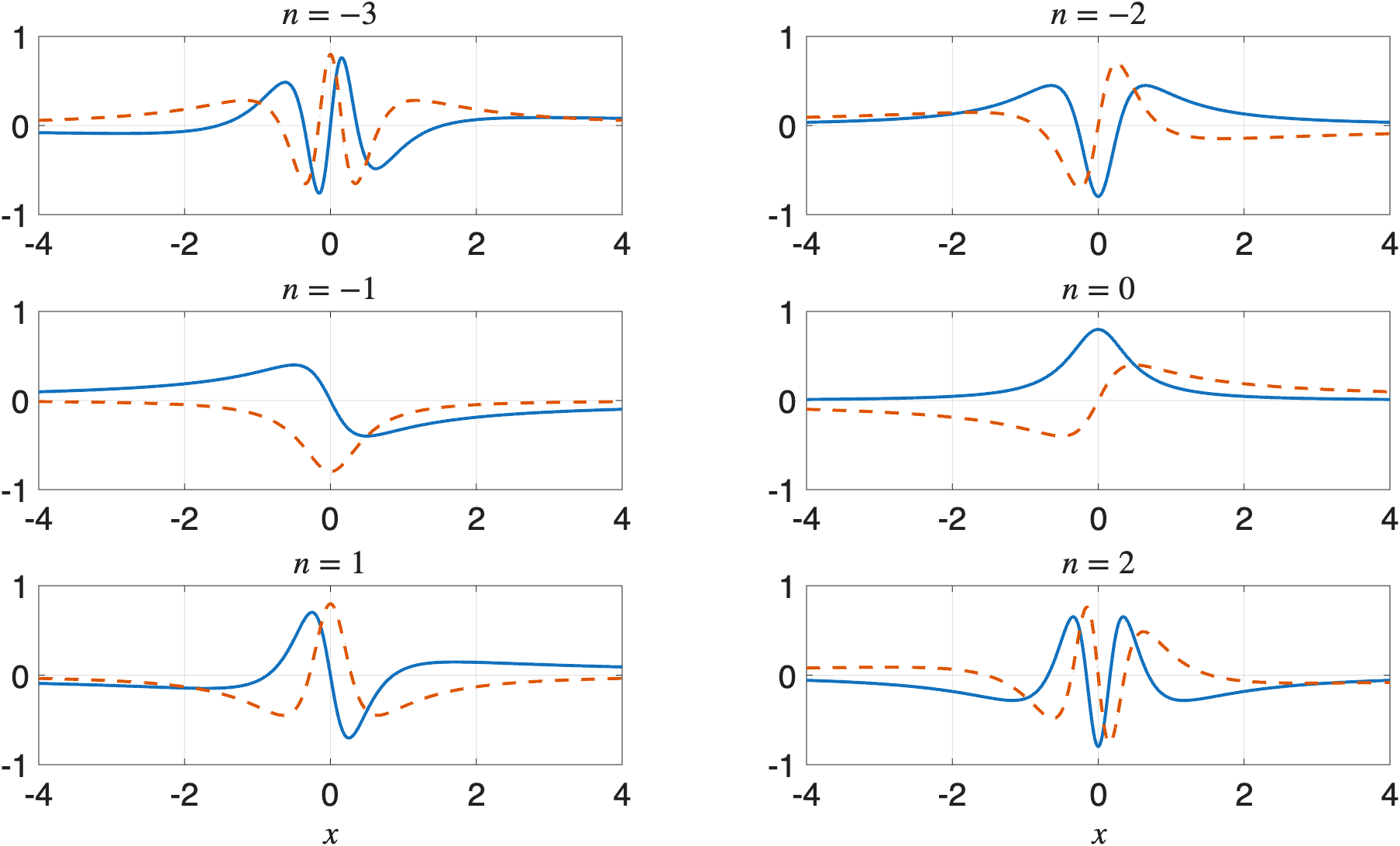}
 \caption{The real and imaginary parts of $\phi_n(x)$ defined in (\ref{Ch_fcn_define}) for $n \in \{-3, \dots, 2 \}$ indicated by blue (solid) and red (dashed) curves, respectively.}
 \label{plot_Ch_fcns}
 \end{figure}

We begin by reviewing some well-known properties of the 
Ch functions.
Define the family of 
Christov rational functions 
 \begin{equation}
 \label{Ch_fcn_define}
   \phi_n(x)=\sqrt{\frac{2}{\pi}}i^n\left(\frac{1+2ix}{1-2ix}\right)^n\frac{1}{1-2ix} , ~~~~~ n \in \ZZ.
 \end{equation} 
  First, observe that $\phi_n(x)$ is well-defined for $x \in \mathbb{R}$ with a pole at $x = -i/2$,  and another $x = i/2$ if $n $ is negative. 
  Several 
  Ch functions are plotted in Fig.~\ref{plot_Ch_fcns}. 
  
 In the literature these functions have also been referred to as Malmquist-Takenaka functions \cite{Iserles2020,Shindin2021}. The original works of Malmquist \cite{Malmquist1925} and Takanaka \cite{Takenaka1926} considered Blaschke products which, when all  zeros coincide, resembles the  functions in (\ref{Ch_fcn_define}). Notably, the zeros (poles) of the original Malmquist-Takanaka functions lie inside (outside) the open unit circle. On the other hand,   functions with the same form as those in (\ref{Ch_fcn_define}) were introduced by Christov to solve  boundary value problems  \cite{Christov1982}. Here the zeros and poles are conjugates of each other and lie on the imaginary axis. Since we are applying  these in spirit of the latter, we adopt the naming convention used in \cite{Boyd1990} and refer to these functions as Christov.

  Below is a list of some properties of these functions.

 \begin{itemize}
 
 \item  {\it Boundedness and decay} 

The 
Ch functions are  uniformly bounded on the real line, that is 
\begin{equation}
|\phi_n(x)| = \sqrt{\frac{2}{\pi}}\frac{1}{\sqrt{1+4x^2}} \leq \sqrt{\frac{2}{\pi}} .
\end{equation}
The algebraic decay rate of the Ch functions  as $x  \rightarrow \pm \infty$ is  linear since 
\begin{equation}
|\phi_n(x)| = \sqrt{\frac{2}{\pi}}\frac{1}{\sqrt{1+4x^2}}  \rightarrow \frac{1}{\sqrt{2 \pi}} \frac{1}{|x|} . 
\end{equation} 

 \item  {\it Oscillations} 
 
 The frequency of oscillations is directly related to the magnitude of the modal value $n$.
  As  $|n| \rightarrow \infty ~( \rightarrow 0)$, the functions become more (less) oscillatory near the origin. Express the rational function that lies on the unit circle by $e^{ i  \theta} = \frac{1 + 2 i x}{1 - 2 i x} $. Then  $\phi_n(x)$ contains a factor of the form $e^{ i n \theta} = \cos (n \theta) + i \sin (n \theta)$ whose angular frequency increases with $|n|$.

 \item {\it Symmetry} 
 
  When $n$ is  even,   the real part of $\phi_n(x)$ is even and the imaginary part is odd. On the other hand, if  $n$ is an odd integer, then the real part of $\phi_n(x)$ is odd and the imaginary part  is even. That is,
\begin{align*}
 n~\text{even:} &~~~~~~~ \Re (\phi_n( - x)) =   \Re (\phi_n( x)), ~~~~ \Im (\phi_n( - x)) =  - \Im (\phi_n( x)) ,  \\
  n~\text{odd:} &~~~~~~~ \Re (\phi_n( - x)) =  - \Re (\phi_n( x)), ~~~~ \Im (\phi_n( - x)) =   \Im (\phi_n( x))  .
 \end{align*}
This symmetry is evident in Fig.~\ref{plot_Ch_fcns}.

  \item {\it Orthogonality} 
  
Define the complex $L^2(\mathbb{R})$ inner product
\begin{equation}
\label{define_inner_prod}
\langle f,g \rangle =  \int_{-\infty}^{\infty}f^*(x)g(x)dx ,
\end{equation}
where $^*$ denotes  complex conjugation. Then for distinct modes, the Ch functions are mutually 
orthonormal and complete in $L^2(\RR)$ \cite{Iserles2020,Higgins1977}, that is 
\begin{equation}
\label{ortho_define}
\langle \phi_m,\phi_n \rangle = \int_{-\infty}^{\infty}\phi^*_m(x)\phi_n(x)dx= \delta_{mn} ,
\end{equation}
where  $ \delta_{mn}$ denotes the Kronecker delta function. The coefficients  in (\ref{Ch_fcn_define}) are chosen to also ensure unit norm, i.e. $|| \phi_n || = 1$, for each $n$. We also note that since the Ch functions are orthonormal and complete  in $L^2(\RR)$ 
they form a basis of $L^2(\RR)$.

\item  {\it Derivative recurrence relation} 

The Ch functions satisfy the following skew-symmetric recurrence relation \cite{Christov1982,Iserles2020} 
\begin{equation}
\label{deriv_recurrence}
\frac{d}{dx}\phi_n(x)=-n\phi_{n-1}(x)+i(2n+1)\phi_n(x)+(n+1)\phi_{n+1}(x).
\end{equation}
This formula naturally leads to tridiagonal differentiation matrices of the first derivative.
Contrast this with Fourier methods which are diagonal, i.e. $\frac{d}{dx} \varphi_n(x) = i k n \varphi_n(x) $ where $\varphi_n(x) = e^{ i k n x}$.
Higher-order derivatives can be obtained by differentiating this formula. 

\item  {\it Hilbert transform eigenfunction} 

The Ch functions are eigenfunctions of the Hilbert transform.
The Hilbert transform is defined by 
$$H[u](x)=\frac{1}{\pi} \dashint_{-\infty}^{\infty}\frac{u(y)}{x-y}dy , $$
where the integral is defined in the Cauchy principal values sense
 \begin{equation*}
 \label{Cauchy_PV}
 \dashint_{-\infty}^{\infty}f(x)dx=\underset{\epsilon \to 0^+}{\lim} \left[\int_{-\infty}^{b-\epsilon}f(x)dx \hspace{0.05in}+ \hspace{0.05in} \int_{b+\epsilon}^{\infty}f(x)dx \right] .
\end{equation*}
The Ch functions satisfy the eigenvalue problem
\begin{equation}
\label{Ch_hilbert}
H[\phi_n](x) = - i\hspace{0.05in}\text{sgn}(n)\phi_n(x) ~~{\rm where}~~ \text{sgn}(n)= \begin{cases} 
      1 & \text{if} \hspace{0.3in} n \geq 0 \\
      -1 & \text{if} \hspace{0.3in} n<0 \\
   \end{cases} .
\end{equation}
A natural application of the Ch functions is the approximation of solutions to the Benjamin-Ono equation which includes the Hilbert transform 

 \end{itemize}

\noindent
{\bf Relationship to Fourier series}\\

Throughout this work we consider expanding a function in terms of the  Ch functions, that is
\begin{equation}
\label{Ch_expansion}
 f(x) = \sum_{n = - \infty}^{\infty} \Check{f}_n\phi_n(x),\hspace{0.1in} \text{where} \hspace{0.1in} \Check{f}_n= \langle \phi_n , f  \rangle = \int_{-\infty}^{\infty} \phi^*_n(x) f(x)  ~ dx ,
 \end{equation}
for the Ch functions given in (\ref{Ch_fcn_define}). This expansion is unique and 
the coefficients are obtained by projecting on an arbitrary mode and exploiting the orthogonality (\ref{ortho_define}).
Through a natural change of variable, this Ch expansion can be related to the Fourier series of a periodic function. This is useful because it  allows one to use the Fast Fourier transform (FFT) to compute the Ch coefficients in (\ref{Ch_expansion}). 

Start from the coordinate transformation 
\begin{equation}
\label{x_theta_transform}
e^{i\theta}=\frac{1+2ix}{1-2ix} ~~~  \Leftrightarrow ~~~ x=\frac{1}{2}\tan\left( \frac{\theta}{2} \right)  ,
\end{equation}
which defines a map from the real line $x \in (- \infty, \infty)$ to the finite interval $\theta \in (- \pi , \pi)$.
Applying this change of variable to the Ch functions (\ref{Ch_fcn_define}) yields
\begin{equation*}
 \phi_n\left( \frac{1}{2}\tan\left( \frac{\theta}{2} \right) \right)=\sqrt{\frac{2}{\pi}}i^ne^{i n \theta}\frac{1}{1-i\tan \left(\frac{\theta}{2} \right)} .
 \end{equation*}
 Next, observe that 
$$\frac{1}{1-i\tan \left(\frac{\theta}{2} \right)} = \frac{\cos(\theta/2)}{\cos(\theta/2)-i\sin(\theta/2)}= e^{i \frac{\theta}{2} }\cos\left( \frac{\theta}{2}  \right) , $$
so that 
 \begin{equation}
 \label{Ch_theta} 
 \phi_n(\theta)=\sqrt{\frac{2}{\pi}}i^n e^{i\theta \left(n+\frac{1}{2} \right)}\cos\left(\frac{\theta}{2}\right).
 \end{equation}

  Note that the Ch function in (\ref{Ch_theta}) is $2\pi$-periodic in  $\theta$. Through the change of variable (\ref{x_theta_transform}), the projection integral in (\ref{Ch_expansion}) becomes
 \begin{equation} 
\label{Ch_coeff_theta}
  \Check{f}_n=\int_{-\infty}^{\infty} \phi^*_n(x)  f(x) dx =\frac{(-i)^n}{2\sqrt{2\pi}}\int_{-\pi}^{\pi} f\left(\frac{1}{2}\tan\left(\frac{\theta}{2} \right)\right)\left(1-i\tan \left(\frac{\theta}{2} \right) \right) e^{-i n \theta } d\theta .
  \end{equation}
 Hence, there is a direct relationship between the Ch coefficients and  
  the Fourier coefficients of the (periodic) function  $ f\left(\frac{1}{2}\tan\left(\frac{\theta}{2} \right)\right)\left(1-i\tan \left(\frac{\theta}{2} \right) \right) $ on the interval $(- \pi , \pi )$.  
 
The Ch coefficients of the function $f(x)$ are computed via
\begin{equation}  
\label{Ch_Fourier_abbrev}
 \Check{f}_n =\mathcal{MF}[f(x)]= (-i)^n\sqrt{\frac{\pi}{2}}\mathcal{F}\left[ f \left(\frac{1}{2}\tan \left(\frac{\theta}{2} \right) \right)\left(1-i\tan \left(\frac{\theta}{2} \right) \right) \right] , ~~~~~~ n  \in \mathbb{Z}
  \end{equation}
  and the function is summed (see Eqs.~(\ref{Ch_expansion}) and (\ref{Ch_theta})) by 
 \begin{equation} 
 \label{MIFFT}
f(x) = \mathcal{MF}^{-1}[\Check{f}_n]=\sqrt{\frac{2}{\pi}}\frac{\mathcal{F}^{-1}\left[ i^n\Check{f}_n  \right]}{1-i\tan \left(\frac{\theta}{2} \right)} ,
  \end{equation} 
 where $\mathcal{F}$ and $\mathcal{F}^{-1}$  denote the Fourier and inverse Fourier (semi-discrete) transforms, respectively.  We refer to  $\mathcal{MF}$ and $\mathcal{MF}^{-1}$ as the  {\it modified Fourier} and {\it inverse modified Fourier transforms}, respectively, which we will utilize to compute the Ch coefficients of a given function.  To be clear, we compute the Ch coefficients directly through Fourier transforms.
 Note that the divisor, $1-i\tan \left(\frac{\theta}{2}\right)$, in (\ref{MIFFT}), is included because (\ref{Ch_Fourier_abbrev}) is computing the  Fourier transform of the function $ f\left(\frac{1}{2}\tan\left(\frac{\theta}{2} \right)\right)\left(1-i\tan \left(\frac{\theta}{2} \right) \right) =f\left(x\right)\left(1-2ix\right)$, so one recovers the function $f(x)$ by dividing out the term, $1-i\tan \left(\frac{\theta}{2}\right)$. 
 The discrete versions of the Fourier transform are naturally computed using the fast Fourier transform (FFT) algorithm (see \ref{Spec_coeff_sec}). \\  

\noindent
{\it Parseval relation} \\\\
Similar to Fourier series, a series expansion in Ch functions (\ref{Ch_expansion}) exhibits an equality of the spatial and spectral $L^2$ norms. This fact follows immediately from the orthogonality (\ref{ortho_define}), 
\begin{equation}
\label{parseval}
\int_{-\infty}^{\infty} |f(x)|^2 dx  = \int_{-\infty}^{\infty} \left( \sum_{n} \check{f}_n^* \phi^*_{n}(x) \right) \left( \sum_{n'} \check{f}_{n'} \phi_{n'}(x) \right) dx =  \sum_{n, n'}  \check{f}_n^* \check{f}_{n'}   \int_{-\infty}^{\infty} \phi^*_n(x) \phi_{n'}(x) dx = \sum_{n = - \infty}^{\infty} \left|\check{f}_n \right|^2 .
\end{equation}
This result assumes $L^2(\mathbb{R})$ is finite, which is 
not always  the case (see (\ref{peregrine})).  \\

  \noindent
{\bf Decacy rate of Ch coefficients}\\

In this work we focus on rational functions 
that are square integrable on the real line, i.e. $L^2(\mathbb{R})$. 
Equivalently, we consider rational functions of the form $p(x)/q(x)$, where $p(x)$ and $q(x)$ are polynomials of degree $r$ and $s$, respectively, where $r < s$. We assume  $q(x)$ has no zeros in $\mathbb{R}$.  We can also consider the case $r = s$ when $\frac{p(x)}{q(x)}$ approaches the finite value $\frac{p_{\infty}}{q_{ \infty}}  $ as $x \rightarrow \pm \infty$. This latter class of rational of functions is clearly not in $L^2(\mathbb{R})$, however the shifted function $\frac{p(x)}{q(x)} - \frac{p_{\infty}}{q_{\infty}}  $ decays to zero and  is in $L^2(\mathbb{R})$. Moreover, the functions $\frac{p}{q}$ and $\frac{p}{q}$ - $\frac{p_\infty}{q_\infty}$  have the same derivatives; an observation exploited in Sec.~\ref{num_pde_sec}. 

An essential ingredient for this spectral method, or any method for that matter, to be  effective is that  the coefficients (see (\ref{Ch_expansion})) decay rapidly, which we discuss next.
In the case of Fourier approximations on the real line, the coefficient decay rate is well-known to be related to the analyticity of the function \cite{Paley1934}. Namely, for functions which can be extended to the complex plane,  Fourier coefficients decay exponentially proportional to distance between the real line and singularities. 
 We  apply this intuition here 
 by locating the poles of the  function in (\ref{Ch_coeff_theta}), but point out that the general decay principles  is still an active area of research \cite{Iserles2020}.

 \begin{figure}
    \centering
   \includegraphics[scale=0.48]{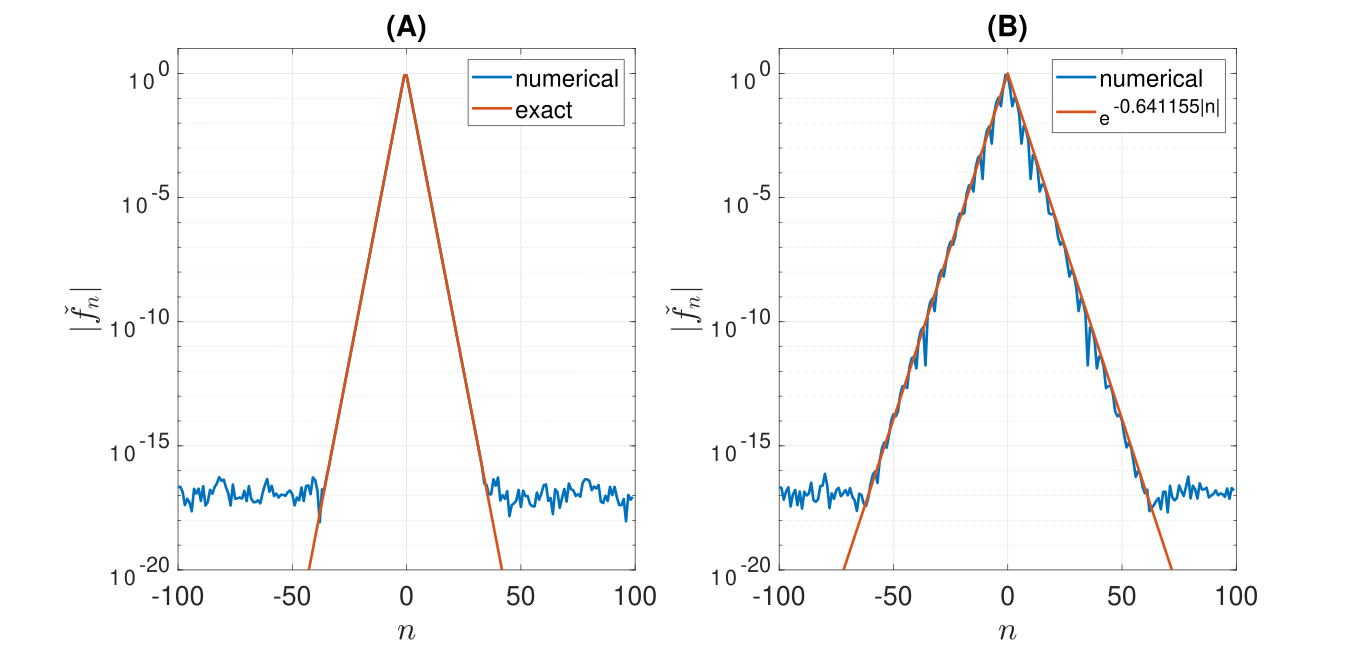}
   \caption{(A) Numerically computed Ch coefficients for the function $f(x) = (1+x^2)^{-1}$ in blue, the exact Ch coefficients  (\ref{rational_expand_exact}) are in red. (B) Numerically computed Ch coefficients for the function $f(x) = (1+x^4)^{-1}$ in blue, the decay rate of Ch coefficients  in red. Notice that 
   the Ch coefficients decay exponentially fast. \label{Ch_coeff_decay_numerical}}
   \end{figure}

As an example, consider the function $f(x) = (1 + x^2 )^{-1}$. 
Under  transformation (\ref{x_theta_transform}), the poles of the function are located at $x = \pm i$ or $\theta = \pm 2 i \tanh^{-1} (2) \approx \pm (\pi + 1.09861 i )$. Since the distance from the real line to the pole is approximately 1.09861, the Fourier-Ch coefficients defined in (\ref{Ch_coeff_theta}) are predicted to decay exponentially with the form $|\check{f}_n| \sim   e^{- 1.09861 |n| }$ as $n \rightarrow \pm \infty$. Actually, $e^{- 1.09861 |n| } \approx 3^{ -|n|}$ which agrees with the exact 
formula for this series
\begin{equation}
\label{rational_expand_exact}
\frac{1}{1 + x^2} = -  \sqrt{2 \pi} \sum_{n = - \infty}^{-1} \frac{i^n}{3^{-n}} \phi_{n}(x) +  \sqrt{2 \pi} \sum_{n = 0}^{\infty} \frac{i^n}{3^{n+1}} \phi_n(x) .\end{equation}
Another example is the function $f(x) = (1 + x^4)^{-1}$ which has poles at $x  = e^{i \frac{\pi}{4} (1 + 2m)}$ for  $m = 0,1,2,3$ or $\theta = 2 \tan^{-1} \left( 2 e^{i \frac{\pi}{4} (1 + 2m)} \right)$. Regardless of which root one takes, the distance from the real line to the imaginary part of the poles is approximately 0.641155. Similar to the previous case, the Ch-Fourier coefficients decay like $|\check{f}_n| \sim  e^{-0.641155 |n| }$ as $n \rightarrow \pm \infty$. Note we predict a slower convergence rate here.

We compare these decay rates with the numerically computed values in  Fig.~\ref{Ch_coeff_decay_numerical}  for the functions $(1+x^2)^{-1}$ and    $(1+x^4)^{-1}$, respectively. 
Importantly, the Ch coefficients of each function decay exponentially fast, reaching round-off error at around $N = 80$ and $N=120$, respectively; a  modest amount of grid points. The rapid decay of the Ch coefficients for slowly decaying rational functions in $L^2(\mathbb{R})$ suggest they could form the basis of an efficient spectral method. The next section considers spectral differentiation using the Ch expansion. This, as with all spectral methods,  benefits from rapidly 
decaying coefficients.

\section{Spectral differentiation}
\label{diff_mat_sec}

Let us now explore differentiation via Ch spectral methods. To begin, consider  a function $f(x)$ expressed in terms of Ch functions as in (\ref{Ch_expansion}) 
whose derivative is 
 \begin{equation*}
 f'(x) = \sum_{n = - \infty}^{\infty} \Check{f}_n \phi'_n(x) .
 \end{equation*}
We wish to express this derivative function in terms of Ch functions itself, that is
 \begin{equation*}
f'(x)   = \sum_{n = - \infty}^{\infty} \Check{f}_n \phi'_n(x)  = \sum_{n = - \infty}^{\infty}  \Check{f}^{(1)}_n \phi_n(x) ,
 \end{equation*}
where $\Check{f}^{(1)}_n$ denotes the derivative function coefficients.
Recall that the derivative of the Ch functions satisfies the recurrence relation (\ref{deriv_recurrence}).
As a result, the Ch coefficients of the derivative can be expressed solely in terms of Ch functions
 \begin{equation*}
 f'(x) = \sum_{n = - \infty}^{\infty} \Check{f}^{(1)}_n \phi_n(x)    = \sum_{n = - \infty}^{\infty} \Check{f}_n \left[ -n\phi_{n-1}(x)+i(2n+1)\phi_n(x)+(n+1)\phi_{n+1}(x) \right] .
 \end{equation*}
After a shift of indices, the Ch coefficients of the derivative  are expressed in terms of the the coefficients of the function itself
\begin{equation}
\label{Ch_deriv_form_series}
 f'(x) = \sum_{n = - \infty}^{\infty} \Check{f}^{(1)}_n \phi_n(x) =   \sum_{n = - \infty}^{\infty}  \left(n \Check{f}_{n-1}  + i(2n + 1) \Check{f}_n - (n +1) \Check{f}_{n+1}  \right) \phi_n(x)  .
\end{equation}
Matching coefficients, the derivative Ch coefficients 
are related to the coefficients of the original function  by
\begin{equation}
\label{deriv_coeff_relate}
\Check{f}^{(1)}_n =  n \Check{f}_{n-1}  + i(2n + 1) \Check{f}_n - (n +1) \Check{f}_{n+1}   .
\end{equation}
Alternatively, the derivative coefficients can be computed via matrix multiplication
\begin{equation}
\label{Ch_diff_mat_relate}
\check{\textbf{f}}^{(1)} = \left( \dots , \check{f}^{(1)}_{-1}, \check{f}^{(1)}_0, \check{f}^{(1)}_1 , \dots \right)^T, ~~~~ \check{\textbf{f}}^{(1)} =\mathbb{D}_1 \check{\textbf{f}} , ~~~~ \check{\textbf{f}} = \left( \dots , \check{f}_{-1}, \check{f}_0, \check{f}_1 , \dots \right)^T ,
\end{equation} 
where $\mathbb{D}_1$ is the banded  matrix
\begin{equation*}
\mathbb{D}_1 =   \left[
\begin{array}{*{10}c}
 \ddots&  \ddots&   &  & &   & & & & \\
 \ddots &  \ddots&  \ddots &  & &   & & & & \\
   &  -3 &  -5i  & 2 & &   & & & & \\
   &   &  -2  & -3i & 1 &   & & & & \\
   &   &    & -1 & -i & 0  & & & & \\
   &   &    &  & 0 & i  & -1 & & & \\
    &   &    &  &  & 1  & 3i & -2 & & \\
    &   &    &  &  &   & 2 & 5i &-3 & \\
     &   &    &  &  &   &  & \ddots & \ddots &\ddots \\
     &   &    &  &  &   &  & & \ddots &\ddots 
\end{array} 
\right] .
\end{equation*}

Notice that this matrix is tridiagonal and skew-Hermitian, that is, $\mathbb{D}_1^\dag = - \mathbb{D}_1$, where $^\dag$ denotes the conjugate transpose. 
The Ch derivative involves self $\{n\}$ and nearest  \{$n-1, n+1$\} modes; this is a direct consequence of recurrence relation (\ref{deriv_recurrence}). On the other hand, many spectral methods often involve completely dense interactions \cite{Trefethen2000}. So, the Ch differentiation matrices are relatively sparse, and as we shall see below, have the ability to converge exponentially fast for appropriate rational functions.  

To approximate derivatives via Ch expansions in practice, we first truncate the series in (\ref{Ch_deriv_form_series}) to a finite range  of $N$ modes: $-N/2, -N/2 + 1 , \dots, N/2-1$, where $N$ is a positive even integer. Then one computes the Ch coefficients of the function $f(x)$ using the discrete version of (\ref{Ch_Fourier_abbrev}), given in (\ref{Ch_discrete_Fourier}). Next, a truncated version of the differentiation matrix in (\ref{Ch_diff_mat_relate}) is applied. The discrete and finite version of the series (\ref{Ch_deriv_form_series}), given in (\ref{inverse_MDFT}), is summed to give the derivative approximation. This process is summarized in \ref{Ch_diff_ap}.

\subsection*{Second-order differentiation matrix} 
\label{higher_order_Ch_diff_sec}

The differentiation matrix for 
the second derivative can be derived in a similar fashion.  Consider a function $f$ expressed in basis of Ch functions, i.e. (\ref{Ch_expansion}), whose  second derivative is clearly
 \begin{equation*}
\label{Ch_expansion_2ndDerivative}
 f''(x) = \sum_{n = - \infty}^{\infty} \Check{f}_n\phi''_n(x) . 
 \end{equation*}
 The goal is to express this function itself in terms of Ch functions, that is
 \begin{equation*}
 \label{Ch_2nd_deriv_form}
 f''(x) = \sum_{n = - \infty}^{\infty} \Check{f}_n \phi''_n(x)   = \sum_{n = - \infty}^{\infty} \Check{f}^{(2)}_n \phi_n(x)   ,
 \end{equation*}
where $\Check{f}^{(2)}_n$ denotes the second derivative coefficients, which we intend to find.
Differentiating the first derivative relation in (\ref{deriv_recurrence}) yields the five-term second derivative relation 
%
 \begin{align}
 \label{Ch_2ndderiv_relation}
\frac{d^2}{dx^2}\phi_n(x) & =-n\frac{d}{dx}\phi_{n-1}(x)+i(2n+1)\frac{d}{dx}\phi_n(x)+(n+1)\frac{d}{dx}\phi_{n+1}(x) \\ \nonumber
& = n(n-1)\phi_{n-2}(x)-4in^2\phi_{n-1}(x)-(2+6n(n+1)) \phi_{n}(x)+4i(n+1)^2\phi_{n+1}(x)+(n+2)(n+1) \phi_{n+2}(x) .
\end{align}
 Note that this involves five nodal values, $\{n-2,n-1,n,n+1,n+2 \}$.
Similar to before, 
shifting the indices gives 
\begin{equation}
\label{Ch_2ndderiv_form_series}
\sum_{n = - \infty}^{\infty} \Check{f}^{(2)}_n \phi_n(x) = 
 \sum_{n = - \infty}^{\infty}  \left[ n(n-1)  \Check{f}_{n-2} + 4in^2 \Check{f}_{n-1}  -(2+6n(n+1))  \Check{f}_{n}  -4i(n+1)^2 \Check{f}_{n+1} + (n+2) (n+1)  \Check{f}_{n+2}   \right] \phi_n(x)  .
\end{equation}
Matching coefficients, the second derivative coefficients are related to the function coefficients via
\begin{equation}
\label{2ndderiv_coeff_relate}
\Check{f}^{(2)}_n = n(n-1)  \Check{f}_{n-2} + 4in^2 \Check{f}_{n-1}  -(2+6n(n+1))  \Check{f}_{n}  -4i(n+1)^2 \Check{f}_{n+1} + (n+2) (n+1)  \Check{f}_{n+2}   .
\end{equation}
The Ch coefficients of the second derivative can also be computed via matrix multiplication
\begin{equation}
\label{Ch_2ndderiv_form}
\check{\textbf{f}}^{(2)} = \left( \dots , \check{f}^{(2)}_{-1}, \check{f}^{(2)}_0, \check{f}^{(2)}_1 , \dots \right)^T, ~~~~ \check{\textbf{f}}^{(2)} =\mathbb{D}_2 \check{\textbf{f}} , ~~~~ \check{\textbf{f}} = \left( \dots , \check{f}_{-1}, \check{f}_0, \check{f}_1 , \dots \right)^T ,
\end{equation} 
where $\mathbb{D}_2$ is the  banded  matrix
$$\mathbb{D}_2 =   \left[
\begin{array}{*{16}c}
 \ddots & \ddots & \ddots &  &  &  &  &  &  &  &  &  &  &  &  &  \\
 \ddots & \ddots & \ddots & \ddots  &  &  &  &  &  &  &  &  &  &  &  & \\
 \ddots & \ddots & \ddots & \ddots  & \ddots &  &  &  &  &  &  &  &  &  &  & \\
  & 30 & 100i & -122  & 64i & 12 &  &  &  &  &  &  &  &  &  & \\
  &  & 20 & 64i  & -74 & 36i & 6 &  &  &  &  &  &  &  &  & \\
  &  &  & 12  & 36i & -38 & 16i & 2 &  &  &  &  &  &  &  & \\
  &  &  &   & 6 & 16i & -14 & -4i & 0  &  &  &  &  &  &  & \\
  &  &  &   &  & 2 & 4i & -2 & 0  & 0 &  &  &  &  &  & \\
  &  &  &   &  &  & 0 & 0 & -2  & -4i & 2 &  &  &  &  & \\
  &  &  &   &  &  &  & 0 & 4i  & -14 & -16i & 6  &  &  &  & \\
  &  &  &   &  &  &  &  & 2  & 16i & -38 & -36i  & 12 &  &  & \\
   &  &  &   &  &  &  &  &   & 6 & 36i & -74  & -64i & 20 &  & \\ 
&  &  &   &  &  &  &  &   &  & 12 & 64i  & -122 & -100i & 30 & \\
&  &  &   &  &  &  &  &   &  &  & \ddots  & \ddots & \ddots & \ddots & \ddots \\ 
&  &  &   &  &  &  &  &   &  &  &   & \ddots & \ddots & \ddots & \ddots \\
&  &  &   &  &  &  &  &   &  &  &   &  & \ddots & \ddots & \ddots
\end{array}
\right] . $$

Notice that this matrix is pentadiagonal and Hermitian ($\mathbb{D}_2^\dag = \mathbb{D}_2$).
Similar to  $\mathbb{D}_1$, this differentiation matrix is banded and fairly sparse; a general theme. Based on  recurrence relation (\ref{deriv_recurrence}),  increasing the order of the derivative increases the bandwidth of the differentiation matrix by one super and one sub diagonal (see \ref{third_Ch_matrix} for $\partial_x^3$). To implement  this in practice, the system in (\ref{Ch_2ndderiv_form}) is truncated to some sufficiently large set of modes and a truncated version of $\mathbb{D}_2$ is applied (see \ref{Ch_diff_ap}). 

A natural question to ask is whether to approximate $\partial_x^2$ by $\mathbb{D}_1^2 = \mathbb{D}_1 \mathbb{D}_1$ (two applications of the first derivative matrix (\ref{Ch_diff_mat_relate})), rather than $\mathbb{D}_2$. When the system is infinite (no truncation), $\mathbb{D}_2 = \mathbb{D}_1^2$ exactly. However, upon truncation the matrices differ at the first and last modes due to different boundary conditions. In our experience, approximation by $\mathbb{D}_2$ is slightly more accurate (by about a factor of 2), but one can use $\mathbb{D}_1^2$ and 
expect similar 
convergence rates.

\begin{figure}
\centering
   \includegraphics[scale=0.4]{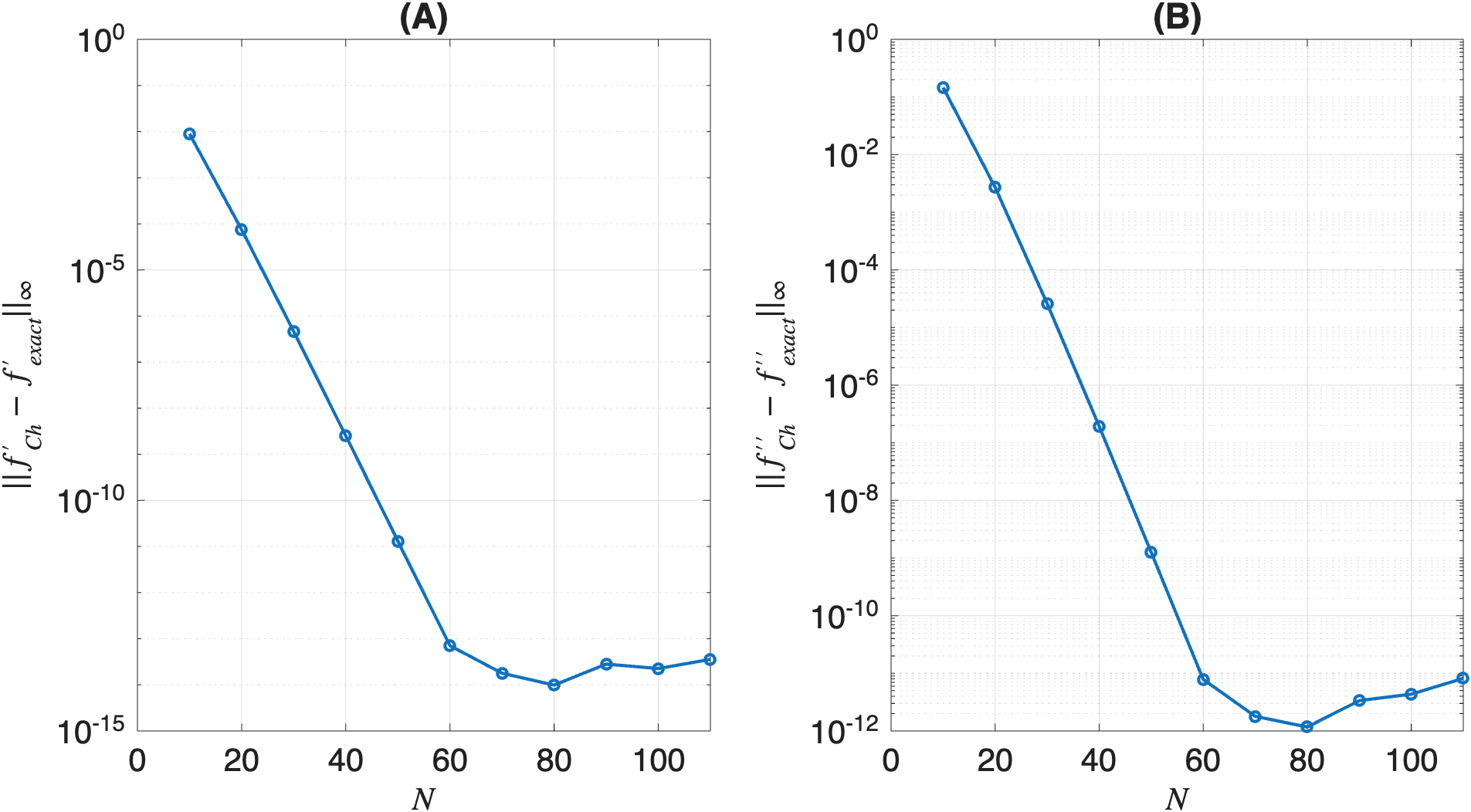}
 \caption{Infinity norm error of the (A) first and (B) second Ch derivative approximations applied to the rational function $f(x) = (1 + x^2)^{-1}$ as a function of the number of Ch modes, $N$. The error approaches zero exponentially fast. All derivatives were computed using the technique described in (\ref{Ch_diff_ap}).  \label{Deriv_converge}}
   \end{figure}

A typical set of convergence results is highlighted in Fig.~\ref{Deriv_converge}. For the rational function $f(x) = 1/(1 +x^2)$, the Ch approximation of derivatives is remarkably effective. The error converges exponentially fast to zero until we reach round-off error effects at around $N = 64$ modes.  We notice with each order of derivative, the error  increases by about one order of magnitude. These rapid convergence results are expected based on the exponentially decay of the Ch coefficients shown in Fig.~\ref{Ch_coeff_decay_numerical}. Given that we are using fast modified Fourier transforms in (\ref{Ch_Fourier_abbrev}) and (\ref{MIFFT}), these calculations are performed quickly.

\section{Numerical approximation of PDEs}
\label{num_pde_sec}

In this section we  develop a method for numerically approximating rational solutions of 
nonlinear  time-{\it dependent} PDEs. A novel explicit split-step time-stepping approach is implemented using the differentiation matrices described in Sec.~\ref{diff_mat_sec}. 

To begin, consider the general  nonlinear initial  value problem
\begin{equation}
\label{general_nonlin_PDE}
u_t=\mathcal{L}u+ \mathcal{M}[u] ,  ~~~~~ u_0 (x) = u(x,t_0) , ~~~~~ (x,t) \in \mathbb{R} \times [t_0, T] , 
\end{equation}
where the linear and nonlinear operators are denoted by $\mathcal{L}$ and $\mathcal{M}$, respectively. 
The linear operator is assumed to be a constant coefficient differential operator of 
the form 
$$ \mathcal{L}= c_0+c_1\frac{\partial}{\partial x} + c_2 \frac{\partial ^2}{\partial x^2} + \dots ~ ,$$
for  scalar coefficients $c_0 , c_1 , \dots$ The nonlinear operator is assumed to be local and depend on some 
product of $u$ and its derivatives.
If both operators in (\ref{general_nonlin_PDE}) are linear ($ \mathcal{M}[u] = M u $) and constant coefficient, then the solution is 
\begin{equation*}
\label{general_nonlin_soln}
u(x,t)  = \exp\left[ (t - t_0) (\mathcal{L} + \mathcal{M}) \right] u_0(x) .
\end{equation*}
 We consider solutions that 
 approach a 
 constant value at the boundaries, i.e. 
 $u(x,t) \rightarrow U_b$ as $x \rightarrow \pm \infty$. Ultimately we have in mind rational solutions where the degree of the denominator polynomial is greater than or equal 
 the degree of the numerator polynomial. The derivatives of the linear operator are approximated by the Ch differentiation matrices (see \ref{Ch_diff_ap}); see example in the next section.

We  now introduce a 
Christov 
approach 
for integrating nonlinear PDEs. 
This method has the benefit of being quite accurate for  rational solutions and straightforward to implement while also
possessing a relatively large region of stability. A thorough derivation of the split-step method and its properties can be found in  \cite{Yang2010,McLachlan2002}. 

Consider splitting (\ref{general_nonlin_PDE}) into two equations 
\begin{equation}
\label{split_step}
\frac{\partial {\bf w}}{\partial t }=\mathcal{L}{\bf w} , ~~~~~~~~ \frac{\partial {\bf v}}{\partial t }= \mathcal{M}[{\bf v}] .
\end{equation}
 The idea of 
 the method is to solve the  equations  in  (\ref{split_step})  separately and then re-combine them in a specific order to approximate the solution. For example, the first-order approximation of (\ref{general_nonlin_soln}) for $t_0 = 0$ is $\exp(t (\mathcal{L} + \mathcal{M})) \approx \exp(t \mathcal{L}) \cdot \exp(t \mathcal{M}).$  That is, solve one equation (\ref{split_step}) and then use its solution as initial condition for the next equation. In general, only when $\mathcal{L}$ and $\mathcal{M}$ commute does this become an equality.
 
We consider the following form 
of a symmetric split-step method for one time-step $\Delta t$ which consists of $s$ solves of the the $\mathcal{L}$ equation and $s+1$ solves of $\mathcal{M}$, in the alternating order 
\begin{equation}
\label{split_step_app}
    {\bf u}(t + \Delta t) \approx \exp\left[  \alpha_{s+1} \Delta t \mathcal{M}  \right]\exp\left[  \beta_{s} \Delta t \mathcal{L}\right]\exp\left[  \alpha_{s} \Delta t \mathcal{M}  \right]\cdots\exp\left[  \beta_1 \Delta t \mathcal{L}\right]\exp\left[  \alpha_1 \Delta t \mathcal{M}  \right]{\bf u}(t) ,
\end{equation}  
where $\alpha_{s+2-i}=\alpha_i$ and $\beta_{s+1-i}=\beta_i$. There have been many formulations of the split-step method; here 
we consider a few fourth-order methods that 
strike a good 
balance between 
accuracy and 
computational cost. The Yoshida (YSH) method \cite{Yoshida1990} is a classic fourth-order symplectic scheme and corresponds to $s=3$; 
we 
refer to this 
as the Ch-YSH method below. More recently, 
several 
 methods have been discovered; 
see \cite{McLachlan2002,McLachlan2021,Blanes2000},  and shown to have better error constants. 
In particular, we examine the $s= 6$
partitioned Runge-Kutta (PRK) and Runge-Kutta-Nystr\"om (RKN) schemes;
 below we 
 refer to these 
 as Ch-PRK and Ch-RKN, respectively. The coefficients for all methods used in this paper can be found in
(\ref{four_coeff_sec}). We note that each time-step of the Ch-PRK and Ch-RKN methods requires 13 substeps, while 
Ch-YSH only requires  7. A comparison of the schemes performances is examined  in more detail below. 


%

 We observe 
 that the overall accuracy of the method does not depend on the choice of which operator is first. 
That is, if the 
method in (\ref{split_step_app}) instead starts and ends with $\mathcal{L}$, 
the order of the method is 
not 
affected.  
We choose to solve the nonlinear part   $s+1$ times ($\alpha_n$ coefficients) and the linear part $s$ times ($\beta_n$ coefficients) because 
for the PDEs studied 
below the nonlinear equation is nearly trivial (and cheap) to solve.

Having introduced the spatial derivative approximation and time-integrator, we fill in the details of a split-step method and  apply it to the NLS equation. The NLS equation is a good candidate for study since it is a physically relevant model that supports interesting rational solutions that  model for rogue waves.



\subsection{Split-step approximation of the NLS equation}
\label{Ch4_sec}


\begin{figure}
    \centering
   \includegraphics[scale=0.4]{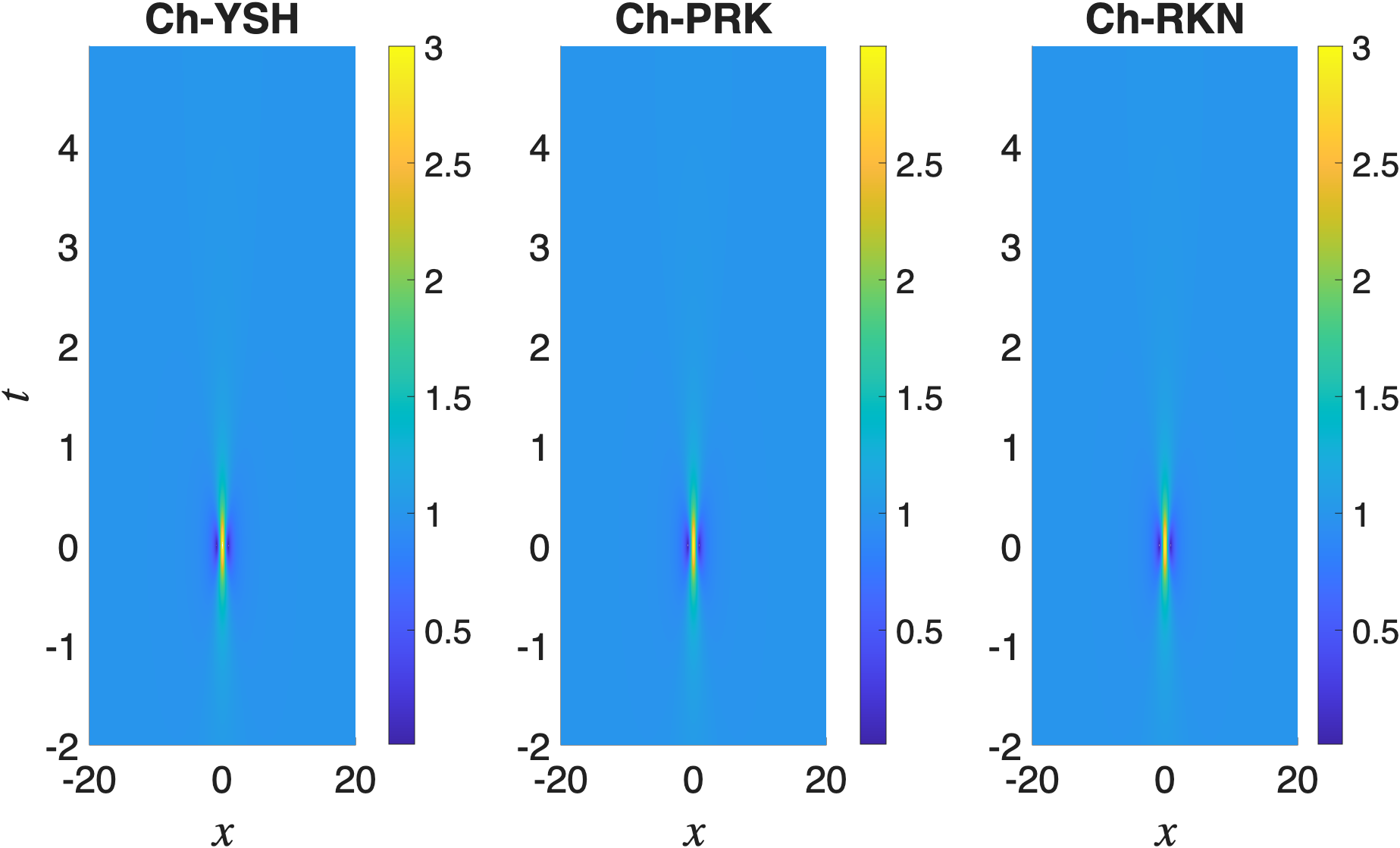}
   \caption{Evolution of $|u(x,t)|$ corresponding to the Peregrine soliton (\ref{peregrine}) in the NLS equation (\ref{nls_eq}) for  $t_0=-2$. Three different Ch split-step  integrators, Ch-YSH, Ch-PRK and Ch-RKN are observed to give similar results.  Computational parameters for each method are
   $N = 320$ Ch modes and time-step $\Delta t=0.001$.   \label{nls_graph} } 
   \end{figure}

The NLS equation is a well-studied integrable equation and a good candidate for gauging the performance of this numerical scheme.
Consider the non-dimensionalized nonlinear Schr\"odinger  equation 
\begin{equation}
    \label{nls_eq}
    iu_t+u_{xx}+2(|u|^2-1)u=0 ,  ~~~~~ u_0 (x) = u(x,t_0) , ~~~~ (x,t) \in \mathbb{R} \times [t_0,T] , 
\end{equation}
with boundary conditions  $u \to 1$ (without loss of generality) as $x \to \pm \infty$. Through the phase transformation $u(x,t) = \psi(x,t) e^{-2 i t }$ this equation can be recast in the more standard form 
$$ i\psi_t+\psi_{xx}+2|\psi|^2 \psi =0 ; $$ the former equation 
is more convenient for our purposes.
Two well-known solutions of the NLS equation are
the (constant) plane wave solution
\begin{equation}
    \label{one}
    u_{c}(x,t)=1 , 
\end{equation}
and Peregrine soliton solution
\begin{equation}
    \label{peregrine}
    u_p(x,t)=\frac{4x^2-16it+16t^2-3}{4x^2+16t^2+1} , 
\end{equation}
both of which are rational functions with unity boundary conditions. Physically, (\ref{one}) corresponds to a nonlinear periodic wave train. 
The Peregrine solution in (\ref{peregrine}) is a model for rogue waves; a depiction of it's dynamics is shown in Fig.~\ref{nls_graph}.
Unlike most solitons that maintain their profile during propagation, the Peregrine soliton is localized in both space {\it and} time. 

Consider solving the NLS equation (\ref{nls_eq}) 
via split equations (\ref{split_step})
where $\mathcal{L} = i \partial_x^2$ and $\mathcal{M}[u] = 2 i \left(|u|^2 - 1 \right)$. Explicitly,  the  split equations 
are
\begin{equation}
\label{NLS_split_step}
\frac{\partial w}{ \partial t } = i \frac{\partial^2 w}{\partial x^2}, ~~~~~~~~ \frac{\partial v}{\partial t} = 2 i \left(|v|^2 - 1 \right) v . 
\end{equation}
The second equation 
is clearly solved by 
\begin{equation}
\label{NLS_nonlinear_solve}
v(x,t_0  + \Delta t) = e^{ 2 i \Delta t \left( |v(x,t_0)|^2  - 1 \right)  } v(x, t_0)  .
\end{equation}

To numerically solve the first equation in (\ref{NLS_split_step}), 
we 
approximate the spatial derivative using an Ch spectral approach. 
Normally, the first step is to expand
\begin{figure}
    \centering
        \includegraphics[scale=0.39]{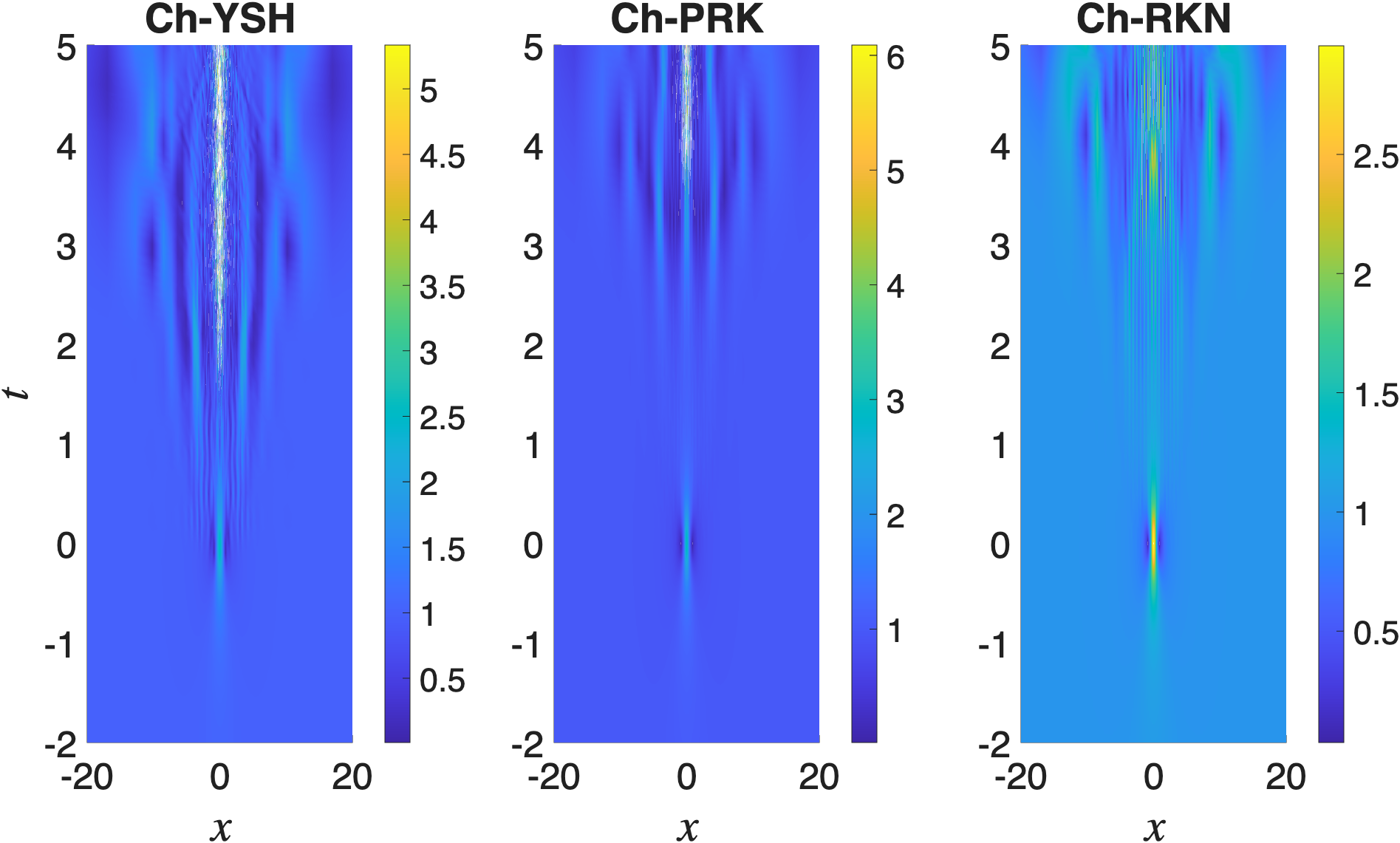}
   \caption{Solution modulus $|u(x,t)|$ using the same   integrators and parameters as Fig.~\ref{nls_graph} except the time-step is now $\Delta t=0.1$. This large time-step reveals  a numerically induced instability in each method. Notice that the instability takes longer to appear for the PRK and RKN methods. 
   \label{unstable_nls_graph} } 
   \end{figure}
$w(x,t)$ in basis of Ch functions (see (\ref{Ch_expansion})). 
However, the solution $w(x,t) \rightarrow 1$ as $|x| \rightarrow \infty$ and is {\it not} in $L^2(\mathbb{R})$;
an expansion in terms of the Ch functions requires the function decay to zero at infinity. As a result, the functions in (\ref{one}) and (\ref{peregrine}) cannot directly be expressed in terms of a Ch expansion.

To get around this, 
first observe the trivial calculus fact  
$$\frac{d f}{d x} = \frac{d}{d x} (f - 1 ) ,$$
for some differentiable function $f(x)$ such that $f(x) \rightarrow 1$ as $|x| \rightarrow \infty$.
Even though both sides of the equation have the same derivative, the function on the right-hand side  has zero boundary conditions and can be directly expanded in a basis of Ch functions. That is, we apply the simple shift 
$\widetilde{w}(x,t) = w(x,t) - 1$ to the linear (first) equation in (\ref{NLS_split_step}), solve, and then return the unit background. 
In this way we can approximate derivatives of functions $u(x)$ with constant {\it and} nonzero boundary conditions, without resorting to a total reformulation of the problem.

After  shifting by one, we expand $\widetilde{w}(x,t)$ as (\ref{Ch_expansion}), and apply 
the differentiation matrix derived in Sec. ~\ref{diff_mat_sec}. 
The system is truncated to $N$ Ch modes and the spectral linear equation
we solve is 
\begin{equation}
\label{lin_dis_nls}
\frac{d \check{\bf{w}}}{d t} = i\mathbb{D}_2 {\check{\bf w}},  
\end{equation}
where $\mathbb{D}_2$ is the differentiation matrix given in (\ref{Ch_2ndderiv_form}) truncated to an $N \times N$ matrix and vector $\check{{\bf w}}$ consists of the Ch coefficients associated with the function $\widetilde{w}(x,t)$.  This is a finite system of ODEs and can be solved exactly by 
\begin{equation}
\label{linear_ss_solve}
\check{{\bf w}}(t_0  + \Delta t ) = e^{i  \Delta t \mathbb{D}_2} \check{{\bf w}}(t_0) . 
\end{equation}
Afterward, the solution in physical space is recovered by computing the modified inverse Fourier transform and adding 1. 

  \begin{figure}
    \centering
   \includegraphics[scale=0.425]{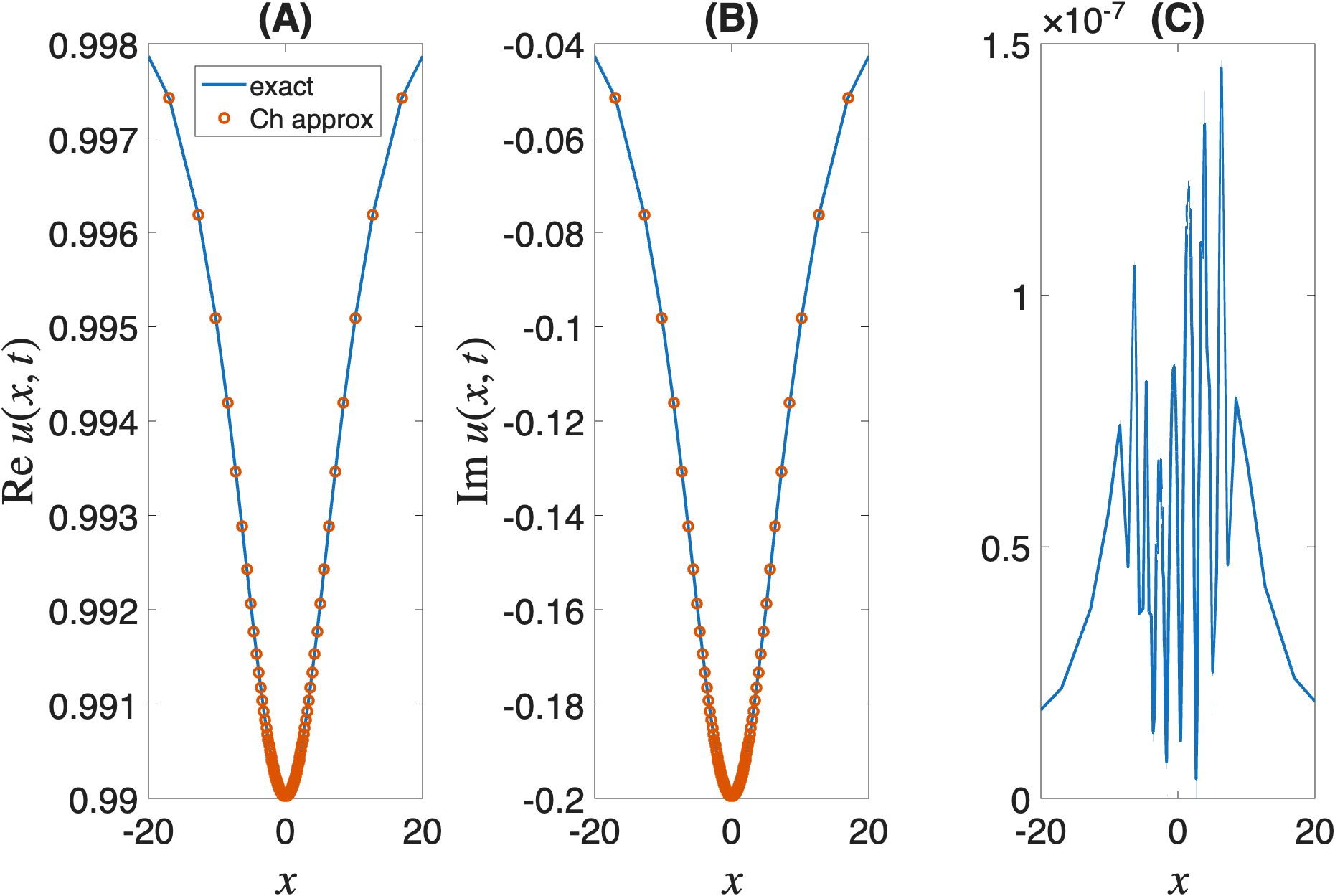}
   \caption{Final time snapshot of the real (A) and imaginary (B) parts of the Ch-RKN approximation for the Peregrine soliton shown in Fig.~\ref{nls_graph}; computational parameters are the same. (C) Absolute difference between Ch-RKN approximation and exact solution at $t = 5$.  \label{nls_snap} } 
   \end{figure}



\begin{figure}
    \centering
   \includegraphics[scale=0.48]{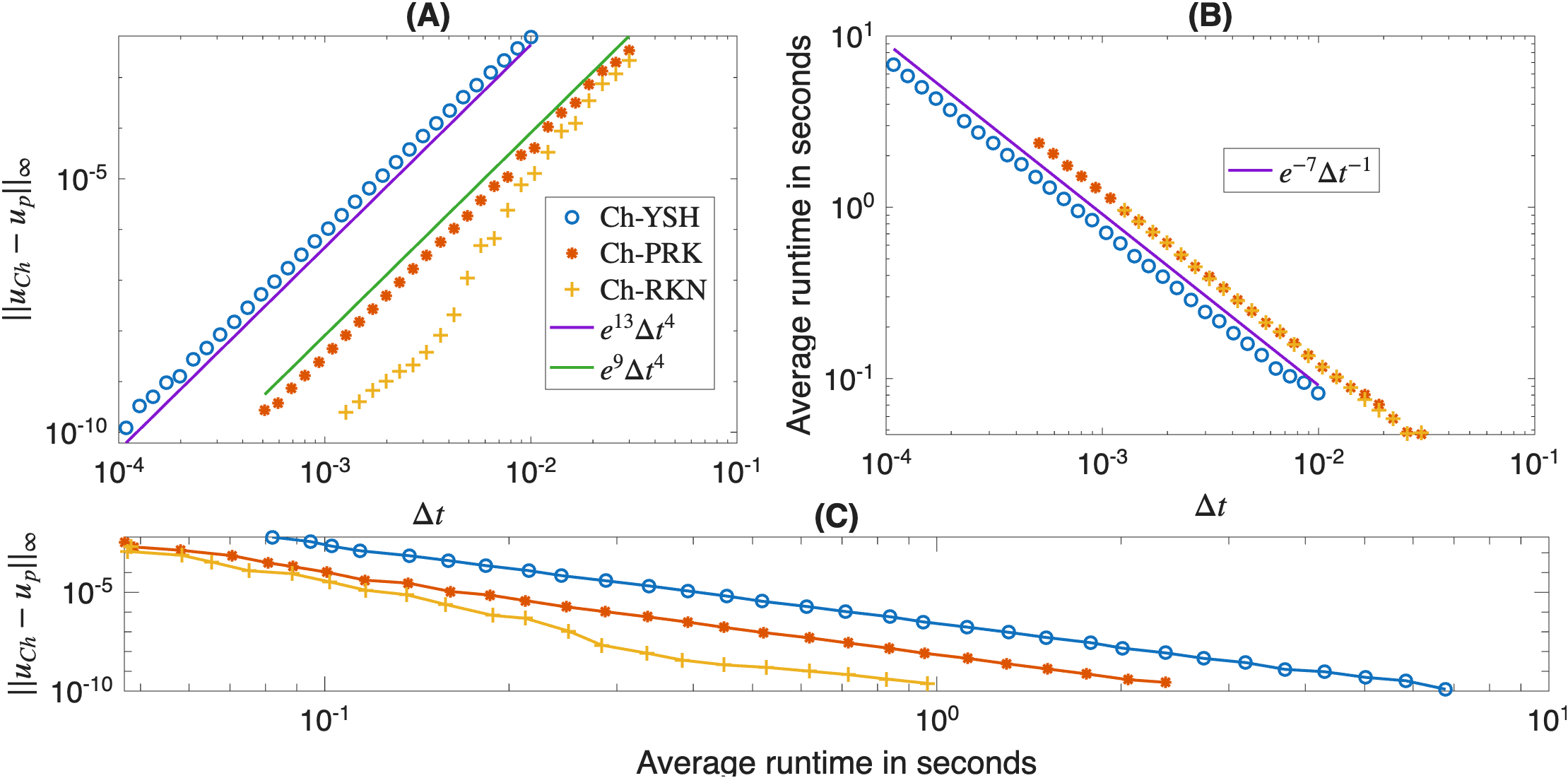}
   \caption{(A) Convergence of 
   Ch split-step approximations for the Peregrine soliton (\ref{peregrine}) solved on the interval $-1 \le t \le 1$. The time-integrators Ch-YSH, Ch-PRK, and Ch-RKN,  defined  in (\ref{four_coeff_sec}), are compared. The error at   $t = 1 $  for $N = 320$  is shown. Overall, each method converges at least fourth-order as  $\Delta t $ decreases. 
   (B) Average CPU runtime as a function of $\Delta t$ and fixed $N$. The runtime  increases linearly with $\Delta t^{-1}$. (C) Comparison of the  average runtime versus the error; the 
   Ch-RKN method performs best. 
   \label{nls_converge}} 
   \end{figure}

\subsection{A convergence study}

Typical results obtained by  solving the NLS equation (\ref{nls_eq})  for the Peregrine soliton solution (\ref{peregrine}) with the Ch-YSH, Ch-PRK, and Ch-RKN  split-step methods are discussed next. The evolution of a typical Peregrine soliton obtained using these methods is shown in Fig.~\ref{nls_graph} over the time interval $[-2,5]$. Notice that the soliton peaks at the origin $(x,t) = (0,0)$. 
For these parameters and time interval, all methods give visually identical results.



A benefit of the split-step method  is the relatively large stability region.
That is, the method yields stable results for quite large values of $\Delta t$. 
A numerically induced instability is shown in Fig. \ref{unstable_nls_graph}. This is the result of taking an apparently too large time-step. The  derivation of a rigorous  stability bound for these solvers  is an open problem. We note that the Ch-PRK and Ch-RKN integrators seem  to postpone the  instability. 

 A comparison between the exact and numerical values is highlighted in Fig.~\ref{nls_snap} at the final time $t = 5$. For a time-step size of $\Delta t = 10^{-3}$, the difference between the exact and numerical approximation is $\mathcal{O}(10^{-7})$ or better. 
 The largest source of 
 error is observed to occur near $x = 0$. 
 We point out that under  transformation (\ref{x_theta_transform}), the spatial points cluster near the origin, even though the points in $\theta$ are evenly spaced.  In practice, we observe  it essential to cluster points near  nontrivial portions of the solution, e.g. rogue wave peak, to achieve maximal accuracy.

Next, the  convergence rates and runtime as a function of $\Delta t$ are 
discussed. In Fig.~\ref{nls_converge}(A) the convergence rates for the three methods 
applied to  the 
 Peregrine solution (\ref{peregrine}) are shown.  For both the Ch-YSH and Ch-PRK integrators, a clear fourth-order convergence rate 
 is observed until other errors become significant. 
 Unexpectedly, while the overall convergence rate of the Ch-RKN methods is fourth-order, there is a section where the error abruptly drops. 
 This is surprising since, in general, this method is fourth-order accurate. 
In previous studies the RKN method was  observed to be 
 well-suited for solving the NLS equation  \cite{Blanes2000}; those results also observed regions of 
 better than expected 
  convergence. A full explanation of this effect is not known at this time. 

 We have observed some apparent aliasing errors for certain $\Delta t $ values. These anomalies can be mitigated by  a spectral padding approach \cite{Canuto2006}. A more detailed analysis of this padding approach is given in (\ref{alias_sec}). 

 The corresponding runtimes are shown in Fig.~\ref{nls_converge}(B). This time encapsulates only  the split-step solver portion of the code. We observe a roughly linear dependence for this range of computational parameters. For all runs considered in Fig.~\ref{nls_converge}, the total CPU time ranged from 0.2-16 seconds. All  runs were performed on a HP Laptop personal computer with an  1.60 GHz 4 Core Intel processor with 8.00 GB of Ram and 15.7 GB of memory.  The data values are the average of over 20 runs to mitigate various CPU-related fluctuations in the runtime. 

 To establish which method gives the best balance between accuracy and efficiency, we plot error vs. average runtime 
 in Fig.~\ref{nls_converge}(C) (combination of panels (A) and (B)), where each data point 
 corresponds to a different value of $\Delta t$. For the Peregrine soliton in this parameter set, the Ch-RKN method provides the highest accuracy for the lowest cost. For example, 
 on the interval 
 $-1 \le t \le 1$ this method is able to reach an error of $10^{-10}$ in about 1.2 seconds on a standard personal laptop. %



Lastly, we do not include any results for the 
plane wave solution 
(\ref{one}) as it is solved {\it exactly} by each Ch split-step method. Examining 
(\ref{NLS_nonlinear_solve}) we see that the nonlinear integrator simply returns the initial plane wave solution, i.e. $v(x, t + \Delta t) = v(x,t)$. For the linear integrator (\ref{lin_dis_nls}), we note that $u_c$ has boundary condition $U_b=1$ and so $\widetilde{w}(x,t) = w(x,t)-1=0$. As a result, all Ch coefficients are zero, $\Check{\bf w} = {\bf 0}$, and  the (trivial) solution to  (\ref{lin_dis_nls}) is zero. Upon returning the boundary condition $U_b = 1$, we recover the exact constant solution, and so on.  

\subsection{Conserved quantities}
\label{conserved_quant_sec}

For sufficiently localized solutions that decay to zero of the NLS equation (\ref{nls_eq}), the following integrals
\begin{equation}
\label{conserved_quant}
P = \int_{-\infty}^{\infty} |u|^2~ dx , ~~~~~~ M = {\rm Im} \int_{-\infty}^{\infty} u^* u_x ~dx , ~~~~~~ H = \int_{-\infty}^{\infty} \left(|u_x|^2 - |u|^4 \right) dx ,
\end{equation}
denoting power, momentum, and Hamiltonian, respectively, are constants of motion. It is desirable that this numerical method 
preserve these invariants.
Since each of the methods 
is symplectic, we expect the Hamiltonian (energy) error to be bounded. 

For  solutions (\ref{one}) and (\ref{peregrine}) with nonzero boundary conditions, the quantities  in (\ref{conserved_quant})  are infinite and so it is not possible to work  with them. However, we can check how the method performs for localized solutions that decay to zero. To approximate the integrals we consider a solution expanded in terms of the Ch basis (\ref{Ch_expansion}). 
One can not approximate the entire integrand by a Ch expansion and integrate since the the Ch functions are not in $L^1(\mathbb{R})$. For example, approximating $|u(x,t)|^2$ by $\sum_n \check{g}_n \phi_n(x)$ yields a power integral that does not converge since $\int_{-\infty}^{\infty} \phi_n(x) ~dx$ does not converge.

The Parseval relation in (\ref{parseval}) is used to verify  power is constant in time. To evaluate 
momentum 
the derivative $u_x$ is approximated through the spectral differentiation matrix in (\ref{Ch_diff_mat_relate}), that is,
\begin{equation}
\label{discrete_momentum}
M  = {\rm Im} ~ \check{\bf u}^\dag (\mathbb{D}_1  \check{\bf u}) , ~~~~~ \check{\bf u} = (\cdots, \check{u}_{-1},  \check{u}_{0}, \check{u}_{1}, \cdots )^T .
\end{equation}
Next, denote $\rho(x,t) = u^2(x,t)$. Then the Hamiltonian is given  by
\begin{equation}
\label{discrete_Hamil}
H  = || \mathbb{D}_1  \check{\bf u}||^2_2  -  ||{\check \rho}||_2^2 , ~~~~~ \check{\bf \rho} = (\cdots, \check{\rho}_{-1},  \check{\rho}_{0}, \check{\rho}_{1}, \cdots )^T .
\end{equation}
Numerically, we truncate these quantities to $N$ modes, where $N \gg 1 $. 

\begin{figure}
    \centering
   \includegraphics[scale=0.45]{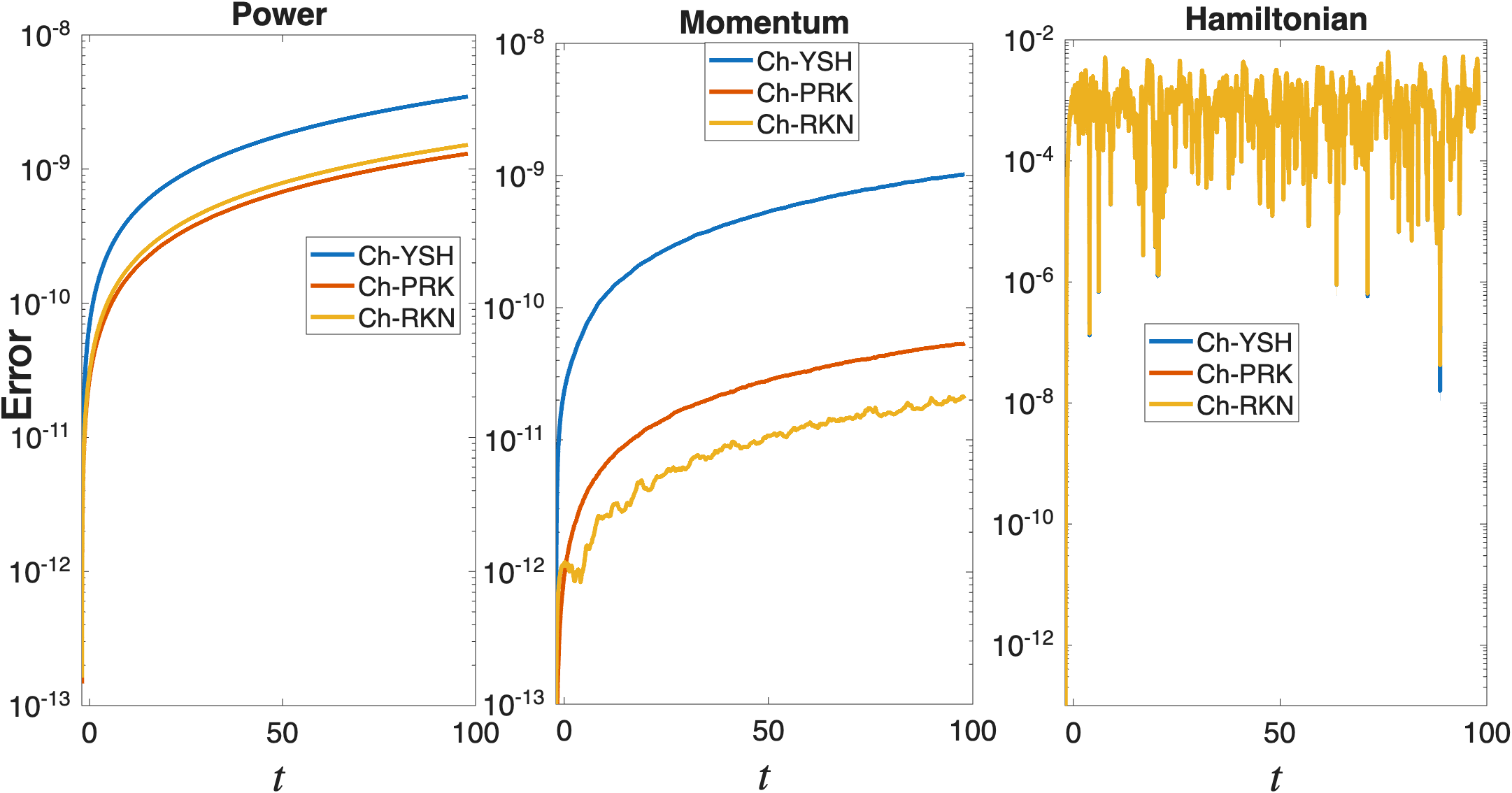}
   \caption{Absolute difference of conserved quantities  (\ref{conserved_quant}) between the initial $t_0 = 0$ and   at $t>0$. The discrete approximation of power, momentum, and Hamiltonian are given in (\ref{parseval}), (\ref{discrete_momentum}), and (\ref{discrete_Hamil}), respectively. 
   The computational parameters used are 
   $N=320$ Ch nodes and $\Delta t=0.001$. 
   \label{con_quant}} 
   \end{figure}

To test the performance of the split-step methods we consider the evolution corresponding to the localized initial condition
$u(x,0)= \frac{1}{1 + x^2} .  $
The error in the power, momentum, and Hamiltonian over the 
time interval $[0,100]$ is shown in Fig.~\ref{con_quant}. We note that all three methods preserve power and momentum quite 
well.  
After some initial 
growth, the error in the  
Hamiltonian remains bounded. 
Other localized initial conditions, e.g. Gaussian, exhibit 
similar results.

Having thoroughly studied the performance of these three splitting methods, we deem the Ch-PRK and Ch-RKN methods to be 
quite effective, 
exhibiting fourth-order convergence or better and preserving conserved quantities. For the remainder of this work we only show the results obtained by Ch-RKN, but note that the 
Ch-PRK method gives similar results. 
Having gained confidence in the method, we next explore  instabilities of the NLS equation with 
nonzero background. 

\section{Instability and rogue waves in the NLS equation}
\label{instab_section}

 In this section we explore two models for rogue waves: the Peregrine soliton solution (\ref{peregrine}) and  modulational instability.  The Peregrine soliton is 
 unstable; our numerical simulations corroborate this. 
Next, we examine generic localized perturbations of the constant background. 
A variety of localized perturbations are found to generate unstable Peregrine-like rogue waves that appear in a recurrent pattern. Both of these results are well-known \cite{Bilman2019,Biondini2016,Kraych2019}  and this numerical scheme should reproduce them.

 \subsection{Instability of the Peregrine soliton}
 
\begin{figure}
    \centering
   \includegraphics[scale=0.425]{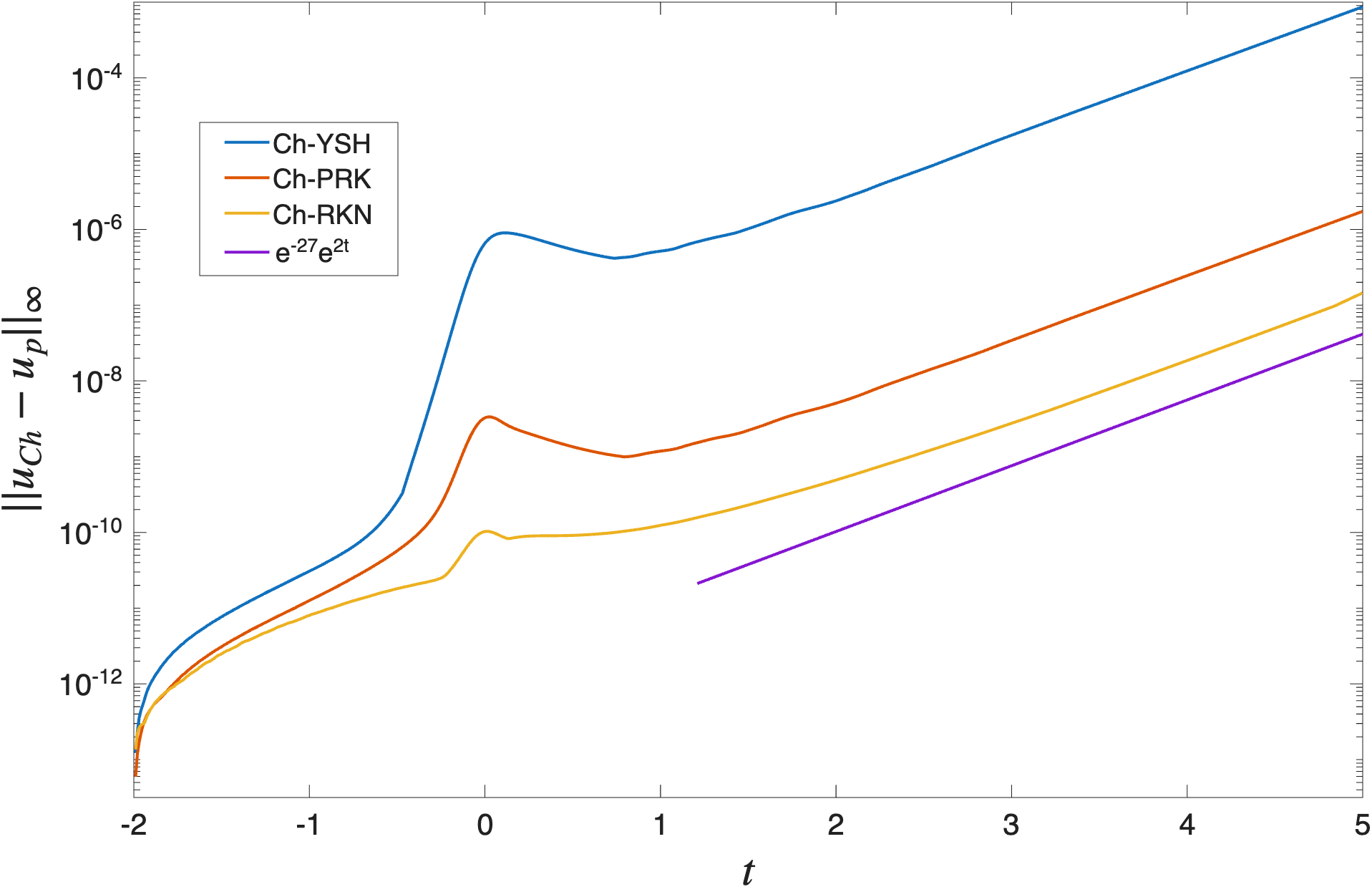}
   \caption{Error in numerical approximation of Peregrine soliton (\ref{peregrine}) over time interval $[-2,5]$ using the Ch split-step methods discussed above for the NLS equation (\ref{nls_eq}) with $t_0=-2$. The computational parameters   are $N = 384$  and $\Delta t = 0.001 $. 
   \label{nls_eVt_growth}} 
   \end{figure}

To begin, we examine the error introduced by  the Ch split-step method when approximating the Peregrine soliton. The infinity-norm error as a function of time is shown in Fig.~\ref{nls_eVt_growth}. 
Importantly, the error is observed to grow exponentially fast for $t > 1$. For linearly stable numerical schemes,  one can bound the error as a function of a term that grows algebraically in $t$ \cite{Cole2024}. 
This on the other hand, is a serious error growth. Decreasing the time step 
is found to postpone, but not eliminate this error growth. 
We have included a set of fits in 
Fig.~\ref{nls_eVt_growth} that illustrate 
the error grows like 
$e^{2t}$. Countless simulations reveal that, 
 regardless of the number of Ch nodes $N$, time-step $\Delta t$, or  
 initial time $t_0$, 
 the growth rate of the error is eventually $\mathcal{O}(e^{2t})$. 
 
 So then, what is the nature of the significant error? Here we attribute it not 
 to a deficiency in our method per se, 
 but rather to the intrinsically unstable nature of the Peregrine soliton and the NLS equation with nonzero boundary conditions. Several recent works  \cite{CuevasMaraver2017,Ablowitz2021,Calini2019}  have highlighted the unstable nature of the Peregrine solution. Moreover, in \cite{Ablowitz2021,Calini2019} the authors precisely predicted a linear instability growth rate of $e^{2t}$. Hence, the method exhibits 
 this expected instability growth rate.

This instability will most likely always be a challenge in simulations of the Peregrine solution. The reason is that numerical approximations always introduce some amount of error, 
 e.g.  round-off. Hence, the moment we try to simulate any 
 solution, regardless of method, we introduce some error. Decreasing the time-step or increasing the number of modes can mitigate these errors, but eventually they will become significant. For context, an initial error of $\delta = 10^{-16}$  will grow of the form $\delta e^{2t}$ and reach a magnitude of $\mathcal{O}(1)$ at around $t \approx 8 \ln 10 \approx 18.4$ time units.


 \subsection{Rogue wave-like structures generated by generic perturbations}

In this section we consider several examples of  localized perturbations of the constant solution (\ref{one}). This would be akin to an ocean wave that is perturbed by a gust of wind or another similar stimuli. 
Starting at an initial time of $t_0=-2$ we perturb the constant solution, 
(\ref{one}), by 
\begin{equation}
\label{perturb_IC}
u_j(x,-2)=1+ \varepsilon f_j(x), ~~~~ j = 1,2,3, 4,
\end{equation}
 for various $\varepsilon$ and a variety of localized functions 
\begin{equation*}
\label{local_fun}
f_1  = \frac{1}{1+x^2} , ~~~~~~
f_2  = \frac{1}{1+x^4} , ~~~~~~
f_3  =e^{-x^2} , ~~~~~~
f_4  = \text{sech}(x) ,
\end{equation*}
 and monitor the evolution. Since we do not have many analytical solutions for this section, we asses the effectiveness of the method by monitoring the decay rate of the Ch coefficients. For all cases below, we require the 
highest-frequency Ch modes  to have a magnitude less than $10^{-3}$ each time-step, that is $|\Check f_{-N/2} | < 10^{-3}$. This was found to ensure results that are visually indistinguishable from more refined ones without imposing a large computational burden.

\begin{figure}
    \centering
   \includegraphics[scale=0.40]{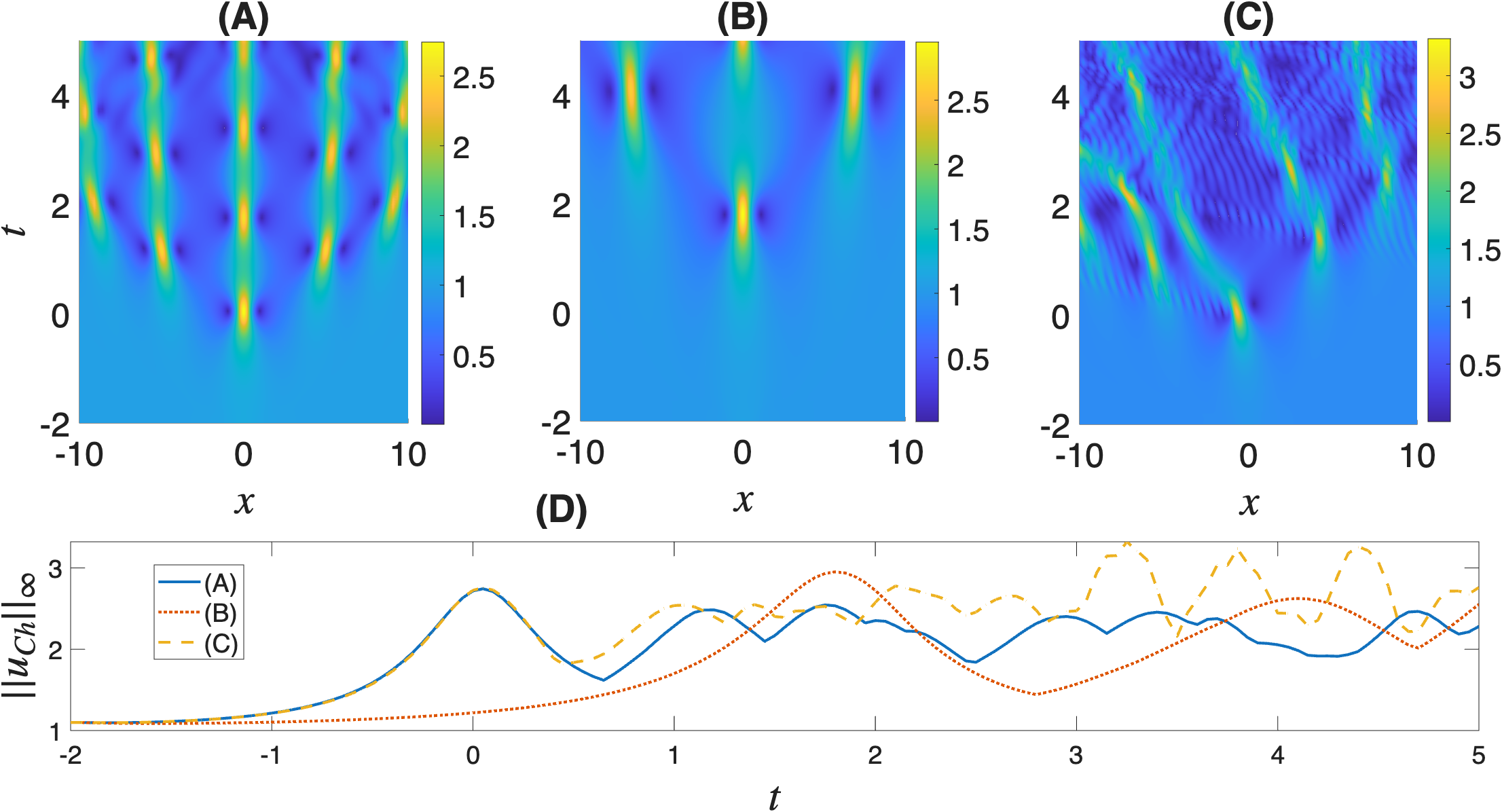}
   \caption{Typical simulations 
   obtained from the 
   perturbed initial condition (\ref{perturb_IC}) with $f_1(x) = \frac{1}{1+x^2}$ and $\varepsilon=0.1$. The corresponding equations are: (A) NLS 
   (\ref{nls_eq}), (B) quadratic-cubic NLS 
   (\ref{gen_nls_eq})  ($p=1$), and (C) higher-order dispersion NLS 
   (\ref{disp_nls_eq})  ($\delta=-0.1$). (D) Maximum amplitude as a function of time. For (A) and (B) the computational parameters used are $N=2000$ and $\Delta t=0.001$ and the computational parameters used for (C) are $N=4000$ and $\Delta t=0.0001$.  
   \label{perturb_movie}} 
   \end{figure}

To begin, we take  $\varepsilon=0.1$. 
For each case considered we observe the formation of a single-hump Peregrine-like structure  
at approximately $t=0$, with a peak height of around 2.7-2.8 times greater than the constant background. A typical evolution is shown in Fig.~\ref{perturb_movie}(A). Notice that the first peak 
is a factor of 27-28 larger than the initial perturbation amplitude ($\varepsilon  = 0.1$). The structure that forms around $t = 0$ bears 
an uncanny resemblance to that of the Peregrine soliton (see Fig.~\ref{nls_graph}). Namely, a localized peak of amplitude nearly 3 
appears and then suddenly disappears. 
Afterward, a two-hump mode emerges and then disappears, followed by a three-hump mode, similar to a modulation instability ``wake'' pattern \cite{Biondini2016,Yuen1978}. Unlike the genuine Peregrine soliton that is localized in time, the 
peaks that form
resemble breathers that are periodic in time.



\begin{figure}
    \centering
   \includegraphics[scale=0.5]{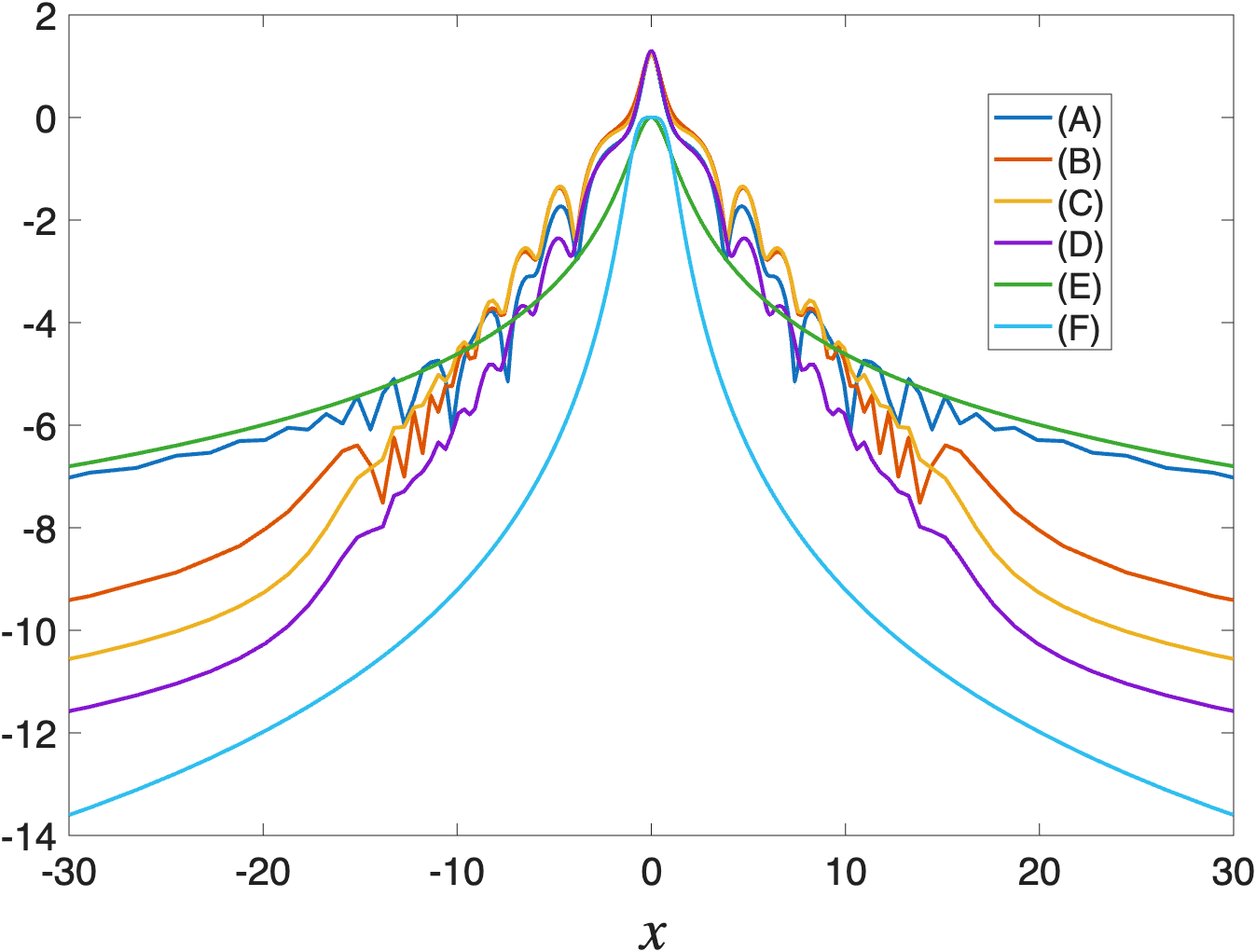}
   \caption{Plots of $\ln  |u(x,0)  - 1|$ obtained by solving the NLS equation (\ref{nls_eq}) 
   with the perturbed initial conditions (\ref{perturb_IC}). These results are obtained from initial perturbations (A) $f_1(x)=(1+x^2)^{-1}$, (B) $f_2(x)=(1+x^4)^{-1}$, (C) $f_3(x)=e^{-x^2}$, (D) $f_4(x)=\text{sech}(x)$.  
   The rational functions $\ln | (1+x^2)^{-1}  |$ and $\ln | (1+x^4)^{-1}  |$ are denoted 
   by curve (E) and (F) respectively; it indicates that the decay rate of all the perturbed approximations are rational. 
   \label{perturb_comp}} 
   \end{figure}

To further establish the Peregrine (rational) nature of this peak, in Fig.~\ref{perturb_comp} we show some solution profiles at $t = 0$ (near the peak time)  generated by each initial condition in (\ref{perturb_IC}). In each case 
the overall decay rate of  $|u(x,0) - 1|$  is found to decay rationally. We have subtracted off the background to isolate the localized part of the solution. Furthermore,  when we apply a variety of rational fits, we consistently find the tails 
decays 
rationally, and at most $\mathcal{O}\left(\frac{1}{x^4} \right)$ as they approach the constant background. 
We conclude that the first peak is nearly the Peregrine soliton at $t = 0$, a universal phenomena in  focusing NLS equations \cite{Tikan2017}.



\begin{figure}
    \centering
   \includegraphics[scale=0.45]{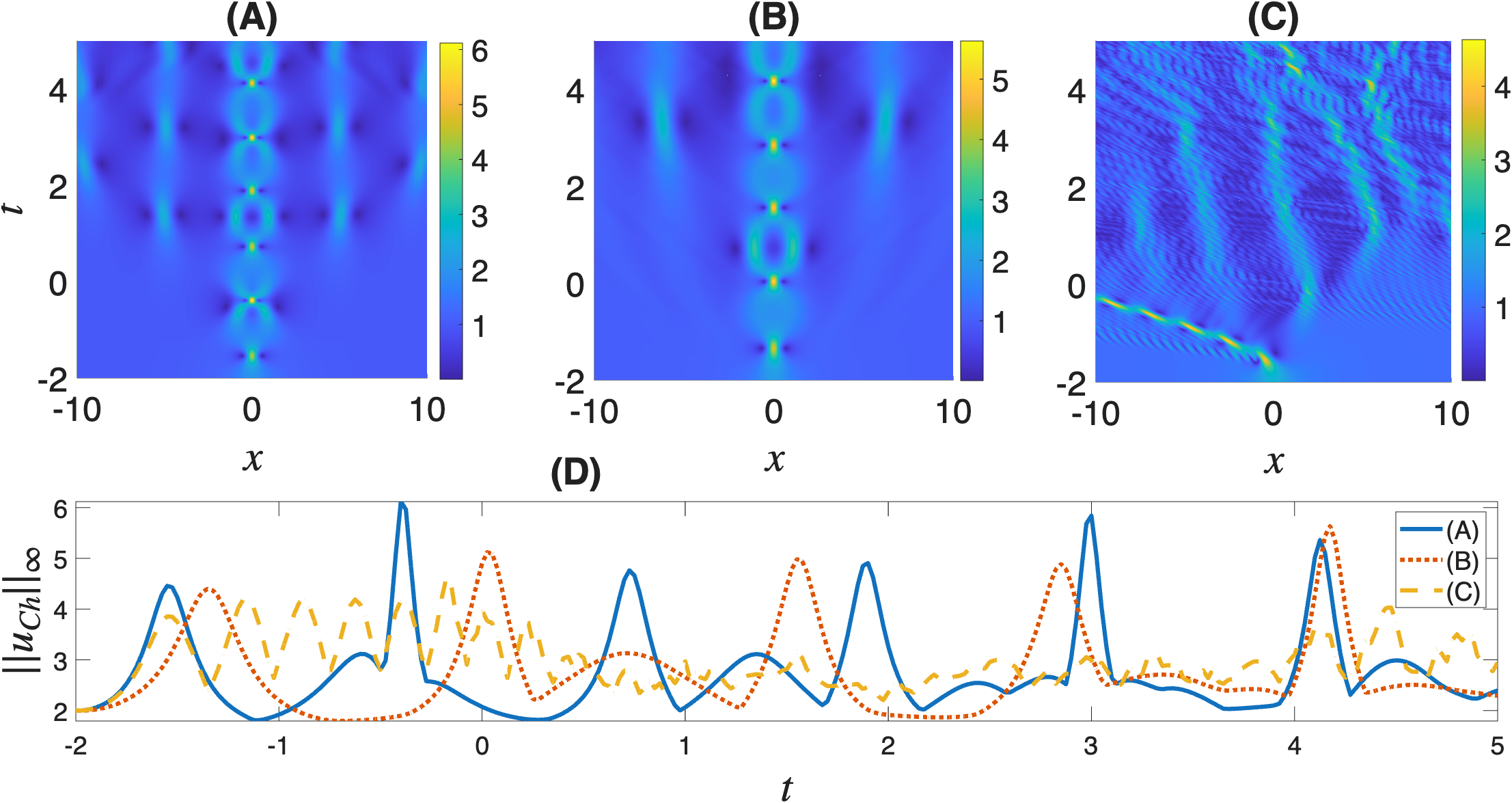}
   \caption{Evolutions for the
   perturbed initial condition (\ref{perturb_IC}) with $f_1(x) = \frac{1}{1+x^2}$ and $\varepsilon=1$. All equations and parameters are the same as Fig.~\ref{perturb_movie}, the only difference is the initial data. 
   \label{ep1_perturb_movie}} 
   \end{figure}

A second case we consider is initial data (\ref{perturb_IC}) with  $\varepsilon=1$, a not small perturbation of the background. 
Here we 
observe a 
single-hump Peregrine-like structure that  appears earlier 
($t \approx -1.75$)
is more localized, and 
has larger amplitude relative to the $\varepsilon = 0.1$ case. 
A breather-like (localized in space, periodic in time) evolution is shown in Fig.~\ref{ep1_perturb_movie}(A). The single-hump state is observed to reoccur in a regular fashion and is reminiscent of the general second-order rogue wave found in \cite{Yang2012}. 
This evolution resembles 
modulational instability on periodic finite domains  dominated by first harmonic modes \cite{Yuen1978}. 
We also note that 
the overall decay rate of  $|u(x,-1.75) - 1|$ at the initial peak is is found to decay rationally, analogous to that observed in Fig.~\ref{perturb_comp}. 




\begin{figure}
    \centering
   \includegraphics[scale=0.40]{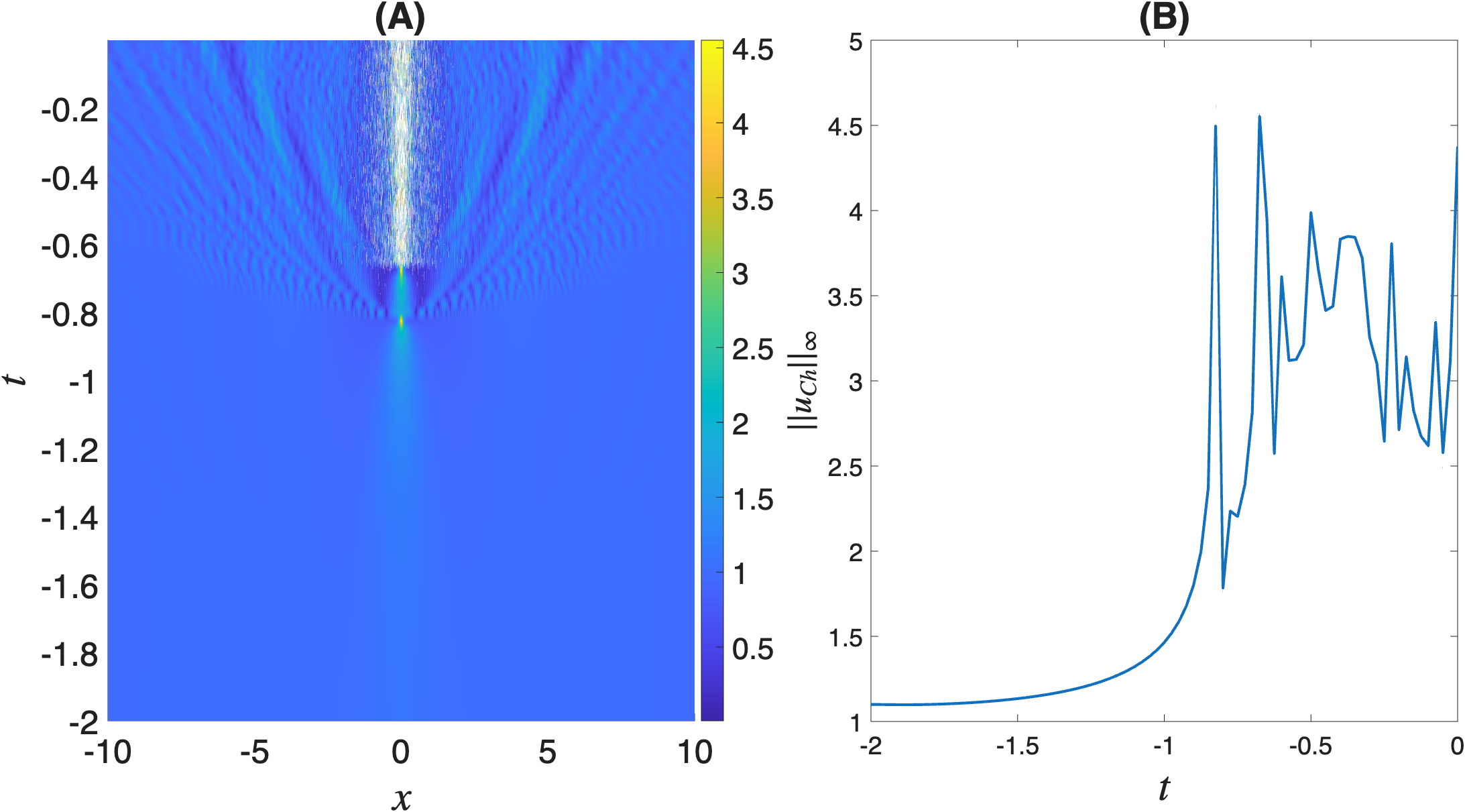}
   \caption{Typical results for the cubic-quintic ($p = 2$) NLS equation (\ref{gen_nls_eq}) obtained by 
   initial condition (\ref{perturb_IC}) with $f_1(x) = \frac{1}{1+x^2}$ and $\varepsilon=0.1$. (A) Evolution of the magnitude $|u(x,t)|$. 
   (B) Maximum amplitude as a function of time; a blow-up appears near $t = -0.8$. The computational parameters used are $N=2000$ and $\Delta t=0.001$. 
   \label{cq_movie}} 
   \end{figure}


\section{
Dynamics in non-integrable NLS-type equations}

The NLS equation is an integrable (exactly solvable) equation which has been thoroughly studied in the literature.
In this section we 
explore the behavior 
of several 
non-integrable extensions of the NLS equation. Namely, we study the formation of rogue wave-like structures in the presence of 
generalized power-type nonlinearity and higher-order dispersion. 
This numerical method is able to shed light on PDEs for which analytical theories, e.g. the inverse scattering transform, are challenging to develop.

\subsection{NLS equation with generalized nonlinearity}

Consider a 
nonlinear Schr\"odinger  equation with a combination of 
power-law type nonlinearities 
\begin{equation}
    \label{gen_nls_eq}
    iu_t+u_{xx}-2|u|^pu+2|u|^{2p}u=0 , 
\end{equation}
where $p$ is a natural number. 
Solutions to \ref{gen_nls_eq}, especially those with smaller values of $p$, appear in various branches of contemporary physics \cite{Mak1990,Raj1982}. For instance, in nonlinear optics the case of $p=2$ appears as a 
model of saturable nonlinearity 
\cite{Caplan2012,bolshinskii1986,mihalache1988,zhou1992}. Intuitively, the nonlinear terms 
compete against each other, focusing $|u|^{2p}$ vs. defocusing $|u|^{p}$, with the higher order focusing  term  dominating at large power (mass). We note for $p=1$, an exact stationary 
bright  
solitary wave solution to (\ref{gen_nls_eq}) is given by \cite{Hayata1995}
\begin{equation}
    \label{bright_sol}
    u_{b}(x,t)=\frac{12}{9+4x^2} .
\end{equation}

The split-step procedure described in (\ref{split_step_app}) can  be applied here. 
We consider a splitting similar to (\ref{NLS_split_step}), with the linear part being identical.
To accommodate the new nonlinearity, we take 
$\mathcal{M}[v] = 2 i \left(|v|^{2p} - |v|^p \right)$. As a result, the solution of the nonlinear part is given by
\begin{equation}
\label{gen_NLS_nonlinear_solve}
v(x,t_0  + \Delta t) = e^{ 2 i \Delta t \left( |v(x,t_0)|^{2p}  - |v(x,t_0)|^p \right)  } v(x, t_0)  ,
\end{equation}
cf. (\ref{NLS_nonlinear_solve}). This  modifies the nonlinear part of the algorithm detailed in \ref{IVPs}.  A natural set of constant boundary conditions  are $u \rightarrow U_b = 0, e^{i \phi } $ 
as $x \rightarrow \pm \infty$. For all simulations in this section, we consider the unit 
boundary condition $U_b = 1$. We have observed 
that 
all three of 
methods discussed in Sec.~\ref{num_pde_sec} do an excellent job approximating the bright soliton solution 
(\ref{bright_sol}), but 
since 
it has no time-dependence the results are not so interesting and we omit them.

Motivated by the previous section, we focus on 
generating rogue wave structures via localized perturbations. 
To remain consistent, we 
start at an initial time of $t_0=-2$ and 
perturb the constant solution (\ref{one}) 
 by (\ref{perturb_IC}) for various $\varepsilon$  and a variety of localized functions 
 and monitor the evolution.
 
Beginning with  quadratic-cubic nonlinearity ($p = 1$),
we observe single-hump Peregrine-like structure (algebraic decay) that  peaks at approximately $t=1.7$ with a peak height of around 
2.9-3.1 times greater than the constant background.  A typical evolution is shown in Fig.~\ref{perturb_movie}(B).
Relative to integrable NLS equation (see Sec.~\ref{instab_section}), these results are 
similar except  
the quadratic term in (\ref{gen_nls_eq}) postpones 
the formation of the rogue wave which is also less localized (longer 
spatial scales). 

Next we move on to  cubic-quintic nonlinearity  ($p = 2$) where the quintic term is focusing and the cubic term defocusing. 
For each initial condition 
considered we observe an apparent blow-up of the solution near time $t = -0.8$.  
A typical evolution is shown in Fig.~\ref{cq_movie}(A). As the dominant nonlinear term is now quintic, 
we expect supercritical NLS behavior, including collapse in finite time 
\cite{Fibich2015}. 
The peak amplitude as a function of time is shown in Fig.~\ref{cq_movie}(B). 
After the apparent collapse time, the simulation gives chaotic results indicative of a breakdown in the PDE model and, consequently, our numerical approximation. 

Lastly, we consider stronger perturbations of the background, taking  $\varepsilon=1$ in (\ref{perturb_IC}). These experiments further elucidate the effect of including additional  nonlinear terms. A typical simulation for the quadratic-cubic nonlinear NLS equation is shown in Fig.~\ref{ep1_perturb_movie}(B). Similar to before, a single-hump Peregrine-like (rational) structure is observed to form initially.
Relative to 
weak perturbations ($\varepsilon = 0.1$), the rogue waves that emerge here have larger amplitude, are more localized, and the dynamics are dominated by single-hump modes; there are weaker 
double-humped modes, for example. Similar to before, the quadratic nonlinearity postpones the first rogue wave event, here forming at 
$t \approx -1.45$. We do not include any simulations for cubic-quintic equation 
and $\varepsilon=1$ case as it is extremely similar to the case of $\varepsilon =0.1$ (see Fig.~\ref{cq_movie}), except 
the collapse occurs earlier 
at $t \approx -1.7$.



\subsection{NLS equation with higher-order dispersion}

We next consider the nonlinear Schr\"odinger  equation with a third-order dispersive correction 
\begin{equation}
    \label{disp_nls_eq}
    iu_t+u_{xx}-i\delta u_{xxx}+2(|u|^2-1)u=0 ,  ~~~~~  |\delta| \ll 1  , 
\end{equation}
and nonzero boundary conditions  $u \to 1$ as $x \to \pm \infty$.
Typically, higher-order dispersion terms are  negligibly small in the slowly-varying envelope approximation, central to the NLS model.  However, these terms  become more significant in the context of rapidly-varying envelopes, such as ultrashort pulses \cite{Agrawal2013}.


We  consider a splitting similar to (\ref{NLS_split_step}) with the same nonlinear part, but linear operator  given by  $\mathcal{L} = i \partial_x^2+\delta\partial_x^3$.
Expanding in a basis of Ch functions, the  linear problem is given by spectral  system and solved
\begin{equation}
\label{disp_lin_dis_nls}
\frac{d \check{\bf{w}}}{d t} = i\mathbb{D} {\check{\bf w}}  ~~~~ \Rightarrow ~~~~ \check{\bf{w}}(t + \Delta t) = e^{i\Delta t \mathbb{D} } \check{\bf{w}}(t) , 
\end{equation}
where $\mathbb{D} =\mathbb{D}_2-i \delta \mathbb{D}_3$, with $\mathbb{D}_2$ being the 
differentiation matrix in 
(\ref{Ch_2ndderiv_form}) and $\mathbb{D}_3$ is the 
differentiation matrix from (\ref{third_Ch_matrix}); cf. (\ref{lin_dis_nls}). 
We will 
again consider initial perturbations of the constant background
(\ref{perturb_IC}) and monitor the evolution. 
For each case discussed,  we take $\delta=-0.1$ to model weak third-order dispersion.

The result of a small amplitude 
localized perturbation is shown in Fig.~\ref{perturb_movie}(C). 
We observe a single-hump Peregrine-like structure  peaks at approximately $t=0$, with a peak height of around 2.9-3 times greater than the unit background, similar to the NLS equation (see Fig.~\ref{perturb_movie}(D)). The results  in Fig.~\ref{perturb_movie}(C) are reminiscent of Fig.~\ref{perturb_movie}(A), but the higher order dispersion creates an effective flow in the negative direction. 
The evolution of a  stronger perturbation is highlighted in Fig.~\ref{ep1_perturb_movie}(C). The main novelty here is 
that the stronger initial perturbation creates a larger amplitude rogue wave-like formation that skews to the left. However, while the evolution is no longer periodic in time due to the skew, the pattern of the evolution indicates that it is still breather-like along the path of the localized wave.


\section{Conclusion}
\label{conclude_sec}

This work 
developed a numerical method for approximating rational solutions of 
certain partial differential equations. 
To approximate these functions,  an expansion in terms of the 
Christov  
functions, a family of  rational orthogonal functions, was proposed. The spectral coefficients can decay exponentially fast for rational functions and 
be numerically computed in terms of (fast) Fourier  methods. This approach represents an alternative to Fourier methods, which struggle with 
slowly-decaying functions on the real line. 

Several explicit, fourth-order and symplectic split-step methods were developed, with the partitioned Runge-Kutta and Runge-Kutta-Nystr\"om being superior. 
As a test, rational ``rogue wave'' solutions of the nonlinear Schr\"odinger equation were considered. The method was found to reproduce known dynamics in the integrable case, e.g.  instability of the Peregrine soliton, as well as reveal interesting behavior in non-integrable variants of the model. 
Looking ahead, a natural application of this method is the study of stability for rogue wave PDE solutions, which are often rational in nature \cite{Feng2015,Feng2023}. \\

\noindent
{\bf Acknowledgments}\\

The authors acknowledge motivating and insightful discussions with Sarbarish Chakravarty and  Denis Silantyev.







\appendix

\section{Discrete modified Fourier and inverse Fourier transforms}
\label{Spec_coeff_sec}

%


In this appendix we discuss the practical implementation 
of (\ref{Ch_Fourier_abbrev}) and (\ref{MIFFT}) to compute the Ch coefficients and sum them via the discrete Fourier transform. We refer to these as the modified fast Fourier transform (MFFT) and modified inverse fast Fourier transform (MIFFT), respectively.
 First we discretize $\theta \in [-\pi , \pi)$ by $\theta_j = j h$, for $j = - \frac{N}{2} , - \frac{N}{2} + 1 , \dots, \frac{N}{2}-1$,  where $N$ is an even integer with angular spacing $h=\frac{2\pi}{N}$. There is no point at $j = N/2$ due to the periodic boundary conditions. The discrete Ch (Fourier) coefficients are  given by
  \begin{equation}
  \label{Ch_discrete_Fourier}
  \Check{f}_n =(-i)^n\sqrt{\frac{\pi}{2}}\frac{1}{N} \sum_{j = -N/2}^{N/2 -1} 
   f \left(\frac{1}{2}\tan \left(\frac{\theta_j}{2} \right) \right)\left(1-i\tan \left(\frac{\theta_j}{2} \right) \right)e^{-in\theta_j} , ~~~~ n = - \frac{N}{2}, - \frac{N}{2} + 1 , \dots, \frac{N}{2}-1 .
  \end{equation} 
This is the discrete approximation of  (\ref{Ch_coeff_theta}). 
Summing the Ch coefficients yields the corresponding function
\begin{equation}
\label{inverse_MDFT}
    f_j =  \sqrt{\frac{2}{\pi}}\frac{1}{1-i\tan\left(\frac{\theta_j}{2}\right)}\sum_{n = -N/2}^{N/2 -1} 
  i^n\Check{f}_ne^{in\theta_j} , ~~~~ j = - \frac{N}{2} , \cdots , \frac{N}{2} -1 ,
\end{equation}
where $f_j = f\left(\frac{1}{2}\tan\left(\frac{\theta_j}{2}\right)\right)$. 
This is the discrete approximation of the sum in (\ref{Ch_expansion}).


\section{Third order Christov differentiation matrix}
\label{third_Ch_matrix}

In this appendix we give the Ch differentiation matrix for the third 
derivative. 
The procedure is similar to those considered in Sec.~\ref{diff_mat_sec}.
We seek to approximate the third derivative by
 \begin{equation}
 f'''(x)  = \sum_{n = - \infty}^{\infty} \Check{f}_n \phi'''_n(x)  = \sum_{n = - \infty}^{\infty} \Check{f}^{(3)}_n \phi_n(x)   ,
 \end{equation}
where $\Check{f}^{(3)}_n$ denotes Ch coefficients of the third derivative function expressed in terms of a basis of Ch functions, $\left\{ \phi_n \right\}_{n=-\infty}^{\infty}$.
The key to relating these two series is differentiating the relation in (\ref{Ch_2ndderiv_relation}); equivalently, differentiating (\ref{deriv_recurrence}) twice. 
The recurrence relation (\ref{deriv_recurrence})  yields the formulas
\begin{equation*}
\frac{d}{dx}\phi_{n-2}(x)=-(n-2)\phi_{n-3}(x)+i(2n-3)\phi_{n-2}(x)+(n-1)\phi_{n-1}(x) ,
\end{equation*}
and
\begin{equation*}
\frac{d}{dx}\phi_{n+2}(x)=-(n+2)\phi_{n+1}(x)+i(2n+5)\phi_{n+2}(x)+(n+3)\phi_{n+3}(x).
\end{equation*}
Combining these formula, the third derivative of an Ch function can be expressed as a linear combination of the seven neighboring Ch functions, namely
\begin{align}
 \label{Ch_3rdderiv_relation}
\frac{d^2}{dx^3}\phi_n(x) & =-n(n-1)(n-2)\phi_{n-3}(x) \\ \nonumber
 &+3i(n-1)n(2n-1)\phi_{n-2}(x)+3n(5n^2+1)\phi_{n-1}(x) \\ \nonumber
  &-2i(2n+1)(5n(n+1)+3)\phi_{n}(x)-3(n+1)(5n(n+2)+6)\phi_{n+1}(x) \\ \nonumber
 &+3i(n+1)(n+2)(2n+3)\phi_{n+2}(x)+(n+1)(n+2)(n+3)\phi_{n+3}(x) .
 \end{align}
As a result, the third-order Ch coefficient is defined by 
\begin{align}
\check{f}^{(3)}_n  &= (n-2) (n-1) n \check{f}_{n-3} + 3 i (n-1)n (2n - 1) \check{f}_{n-2} - 3 n (5 (n^2 - 1)+ 6)  \check{f}_{n-1} \\\nonumber
&-2i(2n+1)(5n(n+1)+3) \check{f}_n \\ \nonumber
&+ 3 (n+1)(5(n+1)^2+1) \check{f}_{n+1} + 3 i (n+1) (n+2) (2n + 3)\check{f}_{n+2} -(n+3)(n+2)(n+1) \check{f}_{n+3} .
\end{align}
Notice that the third derivative possesses a band limited expression of the seven nearest Ch modes, $\{n- 3, \dots, n+3  \}$.
Similar to the previous cases, the coefficients of the third derivative can be computed via
\begin{equation}
\label{Ch_3ndderiv_form}
\check{\textbf{f}}^{(3)} = \left( \dots , \check{f}^{(3)}_{-1}, \check{f}^{(3)}_0, \check{f}^{(3)}_1 , \dots \right)^T, ~~~~ \check{\textbf{f}}^{(3)} =\mathbb{D}_3 \check{\textbf{f}} , ~~~~ \check{\textbf{f}} = \left( \dots , \check{f}_{-1}, \check{f}_0, \check{f}_1 , \dots \right)^T ,
\end{equation} 
where $\mathbb{D}_3$ is  the formidable   banded  matrix

\scalebox{0.56}{$\mathbb{D}_3 =   \left[
\begin{array}{*{24}c}
 \ddots & \ddots & \ddots & \ddots &  & &  &  &  &  &  &  &  &  &  &  &  &  &  &  &  &  &  & \\
\ddots & \ddots & \ddots & \ddots & \ddots & &  &  &  &  &  &  &  &  &  &  &  &  &  &  &  &  &  & \\
\ddots & \ddots & \ddots & \ddots & \ddots & \ddots &  &  &  &  &  &  &  &  &  &  &  &  &  &  &  &  &  & \\
\ddots & \ddots & \ddots & \ddots & \ddots & \ddots & \ddots &  &  &  &  &  &  &  &  &  &  &  &  &  &  &  &  & \\
 & -720 & -3672i & 7704 & 8490i & -5166 & -1638i & 210 &  &  &  &  &  &  &  &  &  &  &  &  &  &  &  & \\
 &  & -504 & -2520i & 5166 & 5538i & -3258 & -990i & 120 &  &  &  &  &  &  &  &  &  &  &  &  &  &  & \\
 &  &  & -336 & -1638i & 3258 & 3366i & -1890 & -540i & 60 &  &  &  &  &  &  &  &  &  &  &  &  &  & \\
  &  &  &  & -210 & -990i & 1890 & 1854i & -972 & -252i & 24 &  &  &  &  &  &  &  &  &  &  &  &  & \\
   &  &  &  &  & -120 & -540i & 972 & 882i & -414 & -90i & 6 &  &  &  &  &  &  &  &  &  &  &  & \\
     &  &  &  &  &  & -60 & -252i & 414 & 330i & -126 & -18i & 0 &  &  &  &  &  &  &  &  &  &  & \\
        &  &  &  &  &  &  & -24 & -90i & 126 & 78i & -18 & 0 & 0 &  &  &  &  &  &  &  &  &  & \\
        &  &  &  &  &  &  &  & -6 & -18i & 18 & 6i & 0 & 0 & 0 &  &  &  &  &  &  &  &  & \\
         &  &  &  &  &  &  &  &  & 0 & 0 & 0 & -6i & 18 & 18i & -6 &  &  &  &  &  &  &  & \\
         &  &  &  &  &  &  &  &  &  & 0 & 0 & -18 & -78i & 126 & 90i & -24 &  &  &  &  &  &  & \\
          &  &  &  &  &  &  &  &  &  &  & 0 & 18i & -126 & -330i & 414 & 252i & -60 &  &  &  &  &  & \\
          &  &  &  &  &  &  &  &  &  &  &  & 6 & 90i & -414 & -882i & 972 & 540i & -120 &  &  &  &  & \\
           &  &  &  &  &  &  &  &  &  &  &  &  & 24 & 252i & -972 & -1854 & 1890 & 990i & -210 &  &  &  & \\
            &  &  &  &  &  &  &  &  &  &  &  &  &  & 60 & 540i & -1890 & -3366i & 3258 & 1638i & -336 &  &  & \\
             &  &  &  &  &  &  &  &  &  &  &  &  &  &  & 120 & 990i & -3258 & -5538i & 5166 & 2520i & -504 &  & \\
              &  &  &  &  &  &  &  &  &  &  &  &  &  &  &  & 210 & 1638i & -5166 & -8490i & 7704 & 3672i & -720 & \\
              &  &  &  &  &  &  &  &  &  &  &  &  &  &  &  &  & \ddots & \ddots & \ddots & \ddots & \ddots & \ddots & \ddots \\
              &  &  &  &  &  &  &  &  &  &  &  &  &  &  &  &  &  & \ddots & \ddots & \ddots & \ddots & \ddots & \ddots \\
              &  &  &  &  &  &  &  &  &  &  &  &  &  &  &  &  &  &  & \ddots & \ddots & \ddots & \ddots & \ddots \\
              &  &  &  &  &  &  &  &  &  &  &  &  &  &  &  &  &  &  & & \ddots & \ddots & \ddots & \ddots 
\end{array}
\right] .$}
Notice that this matrix is septadiagonal and skew-symmatric. To implement this approach numerically, this system will be truncated to a sufficiently large number of modes. While the number of bands continues to grow, this is still fairly sparse;  efficient numerical techniques should 
exploit this sparseness.

\section{Christov approximation of derivatives}
\label{Ch_diff_ap}

In this appendix we give the general algorithm for approximating spatial derivatives using the Ch spectral approximation. The semi-discrete motivations are shown in (\ref{Ch_Fourier_abbrev}) and (\ref{MIFFT}).
The discrete modified Fourier transforms are computed using standard FFT algorithms. We abbreviate the modified FFT by MFFT and the modified inverse Fourier transform by MIFFT. Their precise definitions are given in \ref{Spec_coeff_sec}.
 \begin{align*}
\text{Input:} ~~~ N \text{(even)}, \hspace{0.15in} \text{function } f(x), \hspace{0.2in} \text{differentiation matrix } \mathbb{D}_m \\
x_{j + \frac{N}{2} +1 }=\frac{1}{2}\tan\left( \frac{ \pi j }{N}  \right), ~~~~ j = - \frac{N}{2},- \frac{N}{2} +1 ,..., \frac{N}{2} -1  \\
 \textbf{x}=(x_1,x_2,...,x_{N})^T\\
 \textbf{f}=(f_1, f_2. ,...,f_{N})^T, \hspace{0.1in} f_j \approx f(x_j)\\
 \Check{\textbf{f}}=\text{MFFT}[\textbf{f}]\\
 \Check{\textbf{f}}^{(m)}=\mathbb{D}_m\Check{\textbf{f}}\\
\textbf{f}^{(m)}=\text{MIFFT}\left[ \Check{\textbf{f}}^{(m)}\right] \\
\text{Output:} ~~~ \textbf{f}^{(m)} = (f_1^{(m)}, f_2^{(m)} ,...,f_{N}^{(m)})^T
 \end{align*}  

This algorithm gives derivative approximations  at the equally spaced angular points $\theta_j = \frac{2 \pi j}{N} $ for $j = -N/2 , -N/2 +1, \dots, N/2 - 1$. Differentiation matrices for the first two derivatives are shown in (\ref{Ch_diff_mat_relate}) and (\ref{Ch_2ndderiv_form}). The output is the approximation of the $m^{\rm th}$-order derivative evaluated at the grid points, i.e. $f_j^{(m)}  \approx f^{(m)}(x_j)$. Once the Ch coefficients  are obtained, one can evaluate $f(x)$ and its derivatives at arbitrary values of $x$ (not just the spatial grid points)  by summing the truncated Ch series $f^{(m)}(x) \approx\overset{N/2-1}{\underset{n=-N/2}{\Sigma}}\Check{f}^{(m)}_n \phi_n(x), ~ m = 0, 1, 2,$ for the Ch functions given in (\ref{Ch_fcn_define}) at an arbitrary 
value of $x$.

\section{Split-step integration}
\label{IVPs}

All PDEs in this work are solved using an explicit Christov split-step  method described in Sec.~\ref{Ch4_sec}. These  methods use a Ch approximation in space and a fourth-order method in time.  
The algorithm pseudo-code is given below. 


The temporal interval $[t_0,T]$ is discretized into $S+1$ equally spaced time levels $t_l = t_0 + l \Delta t$ for  $l = 0 ,1 , \dots, S$ with time-steps of $\Delta t  = \frac{T-t_0}{S}$. The real line $(- \infty , \infty)$  is discretized by $x_{j + \frac{N}{2} +1 }=\frac{1}{2}\tan\left( \frac{ \pi j }{N}  \right),$ for $ j = - \frac{N}{2},- \frac{N}{2} +1 ,..., \frac{N}{2} -1$ and  
$ {\bf x}=(x_1,x_2,...,x_{N})^T$ for even $N$.  Denote the numerical approximation of the solution by ${\bf u}(t) =(u_1,u_2,...,u_{N})^T $ where $u_j  \approx u(x_j,t )$.  The finite second differentiation matrix $\mathbb{D}_2$ is defined in (\ref{Ch_2ndderiv_form}) and truncated to a matrix of size $ N \times N$. 
The initial condition is given by ${\bf u}_0 = u({\bf x}, t_0)$.

Consider the NLS equation (\ref{nls_eq}) and  splitting (\ref{NLS_split_step}). The solution of the nonlinear equation was given in (\ref{NLS_nonlinear_solve}); notice this is solved in the physical domain. To solve the linear equation, first the solution is shifted, ${\bf u} - 1$, to ensure zero boundary conditions. Next, the MFFT transform (\ref{Ch_discrete_Fourier}) is taken to find the coefficients and the linear system is solved by (\ref{linear_ss_solve}). Afterward, the function is reconstructed by MIFFT transform (\ref{inverse_MDFT}) and the unity boundary conditions are restored. The coefficients in the split-step method given in (\ref{split_step_app}) can be chosen to 
achieve fourth-order accuracy.  

\vspace{0.2 in}

\noindent
{\it One Ch split-step time-step for  the NLS equation}

\begin{align*}
{\bf v}_1 =\exp\left[2i \alpha_1\Delta t  \left(|{\bf u}_j|^2-1\right) \right]{\bf u}_j \\
\Check{{\bf v}}_1 =\text{MFFT}[{\bf v}_1-1]\\
\Check{{\bf w}}_1=\exp\left[i \beta_1\Delta t \mathbb{D}_2 \right]\Check{{\bf v}}_1 \\
{\bf y}_1 =\text{MIFFT}\left[ \Check{{\bf w}}_1 \right] + 1 \\
{\bf v}_2 =\exp\left[2i \alpha_2\Delta t \left(|{\bf y}_1|^2-1\right) \right]{\bf y}_1 \\
\Check{{\bf v}}_2 =\text{MFFT}[ {\bf v}_2-1]\\
\Check{{\bf w}}_2=\exp\left[i\beta_2\Delta t \mathbb{D}_2 \right]\Check{{\bf v}}_2 \\
{\bf y}_2 =\text{MIFFT}\left[ \Check{{\bf w}}_2 \right] + 1 \\
\vdots \\
{\bf v}_s =\exp\left[2i \alpha_s\Delta t \left(|{\bf y}_{s-1}|^2-1\right) \right]{\bf y}_{s-1}\\
\Check{{\bf v}}_s =\text{MFFT}[ {\bf v}_s-1]\\
\Check{{\bf w}}_s =\exp\left[i\beta_s\Delta t \mathbb{D}_2 \right]\Check{{\bf v}}_s \\
{\bf y}_s =\text{MIFFT}\left[ \Check{{\bf w}}_s \right] + 1 \\
{\bf u}_{j+1}=\exp\left[2i \alpha_{s+1}\Delta t \left(|{\bf y}_s|^2-1\right) \right]{\bf y}_s \\
 t_{j+1} =t_j+\Delta t
\end{align*}

\section{Coefficients of fourth-order split step methods}
\label{four_coeff_sec}
The general form of the split-step time-stepping method is given in (\ref{split_step_app}). The three methods we considered can be found in \cite{Yoshida1990,McLachlan2021,Blanes2000}.  We print them below, for convenience. \\

Coefficients of the Yoshida (YSH) fourth-order split-step 
method, where $s=3$, are:
\begin{align*}
\alpha_1=\frac{1}{2}c, \hspace{0.1in} \alpha_2=\frac{1}{2}(1-c),\hspace{0.1in}\alpha_3=\alpha_2,\hspace{0.1in}\alpha_4=\alpha_1\\
    \beta_1=c,\hspace{0.1in}\beta_2=1-2c, \hspace{0.1in}\beta_3=\beta_1 
\end{align*}
where $c=\frac{1}{2-2^{1/3}}$.

Coefficients of a fourth-order partitioned Runge-Kutta method (PRK), where $s=6$, are:
\begin{align*}
\alpha_1=0.0792036964311957, \hspace{0.1in}\alpha_2=0.353172906049774, \hspace{0.1in}\alpha_3=-0.0420650803577195\\
\alpha_4=1-2(\alpha_1+\alpha_2+\alpha_3), \hspace{0.1in}\alpha_5=\alpha_3, \hspace{0.1in}\alpha_6=\alpha_2, \hspace{0.1in}
\alpha_7=\alpha_1\\
\beta_1=0.209515106613362, \hspace{0.1in}\beta_2=-0.143851773179818\\
\beta_3=0.5-(\beta_1+\beta_2), \hspace{0.1in}
\beta_4=\beta_3, \hspace{0.1in}
\beta_5=\beta_2, \hspace{0.1in}
\beta_6=\beta_1
\end{align*}

Coefficients of a fourth-order 
Runge-Kutta-Nystr\"om method (RKN), where $s=6$, are:
\begin{align*}
\alpha_1=0.0829844064174052, \hspace{0.1in}\alpha_2=0.396309801498368, \hspace{0.1in}\alpha_3=-0.0390563049223486\\
\alpha_4=1-2(\alpha_1+\alpha_2+\alpha_3), \hspace{0.1in}\alpha_5=\alpha_3, \hspace{0.1in}\alpha_6=\alpha_2, \hspace{0.1in}\alpha_7=\alpha_1\\
\beta_1=0.245298957184271, \hspace{0.1in}\beta_2=0.604872665711080\\
\beta_3=0.5-(\beta_1+\beta_2), \hspace{0.1in}\beta_4=\beta_3, \hspace{0.1in}\beta_5=\beta_2, \hspace{0.1in}\beta_6=\beta_1
\end{align*}

\section{Aliasing errors}
\label{alias_sec}

In this appendix we address an aliasing-type error that can occur with this method. Aliasing effects are a well-known consequence in certain spectral method approximations of nonlinear PDEs \cite{Canuto2006}. In our case, sometimes when  implementing our time-stepping  algorithms for a particular $\Delta t$, we observe anomalously large errors relative to similar time-step sizes. This is due to  unexpectedly large high-frequency Ch modes. We propose a spectral padding approach to mitigate these effects.

\begin{figure}
    \centering
   \includegraphics[scale=0.45]{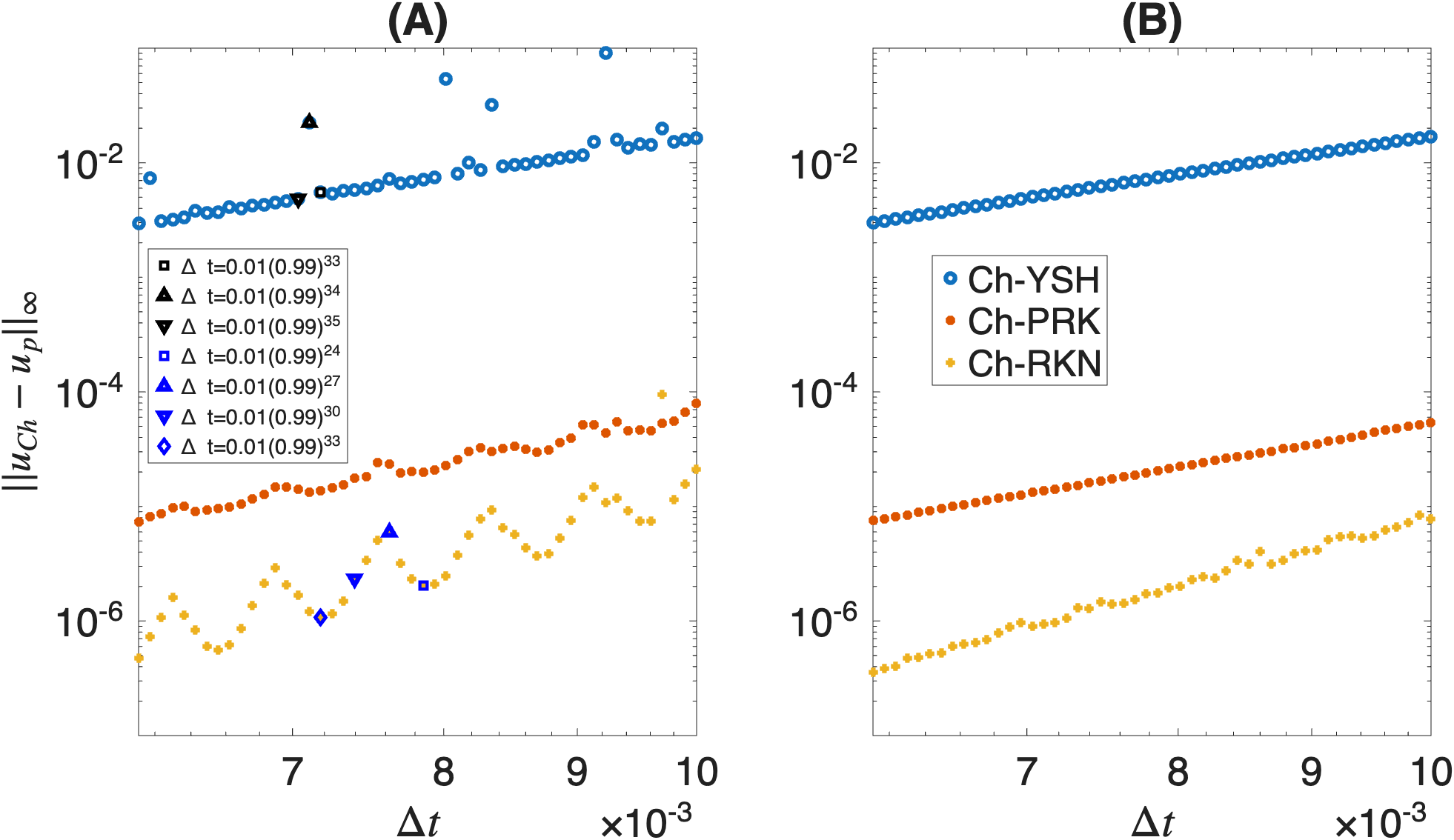}
    \caption{Convergence of the Ch split-step methods 
    for the Peregrine soliton (\ref{peregrine}). Other than the time-step $\Delta t$, and the time interval, $-2 \leq t \leq 2$, all computational parameters are the same as Fig.~\ref{nls_converge}.  
    (A) The numerical approximation  converges as expected except for a few problematic $\Delta t$ values. Specific points where convergence jumps occur are indicated. (B) Convergence rate of the {\it padded} split-step 
    methods defined in  (\ref{padding_algorithm})  with $P = 10$ applied; otherwise, this is the same as (A).  \label{nls_alias}}
   \end{figure}

The error convergence rates for the Ch-YSH, Ch-PRK and Ch-RKN methods applied to the NLS equation (\ref{nls_eq}) are shown in Fig.~\ref{nls_alias} for the Peregrine soliton. The only difference between this graph and that of Fig.~\ref{nls_converge}(A) is that we include  more values of $\Delta t$ and we  compute the equation on the interval $-2 \leq t \leq 2$ as the aliasing errors become more apparent. 
In Fig.~\ref{nls_alias}(A)  we multiply 
the value of 
$\Delta t$ by 
$0.99$ each run 
(equivalently, decrease 
by a factor of $100/99$). 
The overall convergence rate appears fourth-order except for a few problematic $\Delta t$ values. 
The Ch-YSH method has anomaly at 
$\triangle$: $\Delta t=0.01(0.99)^{34}$ where the error is $\OO(10^{-2})$, meanwhile a slightly larger 
value of $\square$: $\Delta t=0.01(0.99)^{33}$ or slightly smaller 
value of $\triangledown$: $\Delta t=0.01(0.99)^{35}$ 
have errors that are $\OO(10^{-3})$. This order of magnitude increase of error is concerning and requires further investigation.   

\begin{figure}
    \centering
   \includegraphics[scale=0.3]{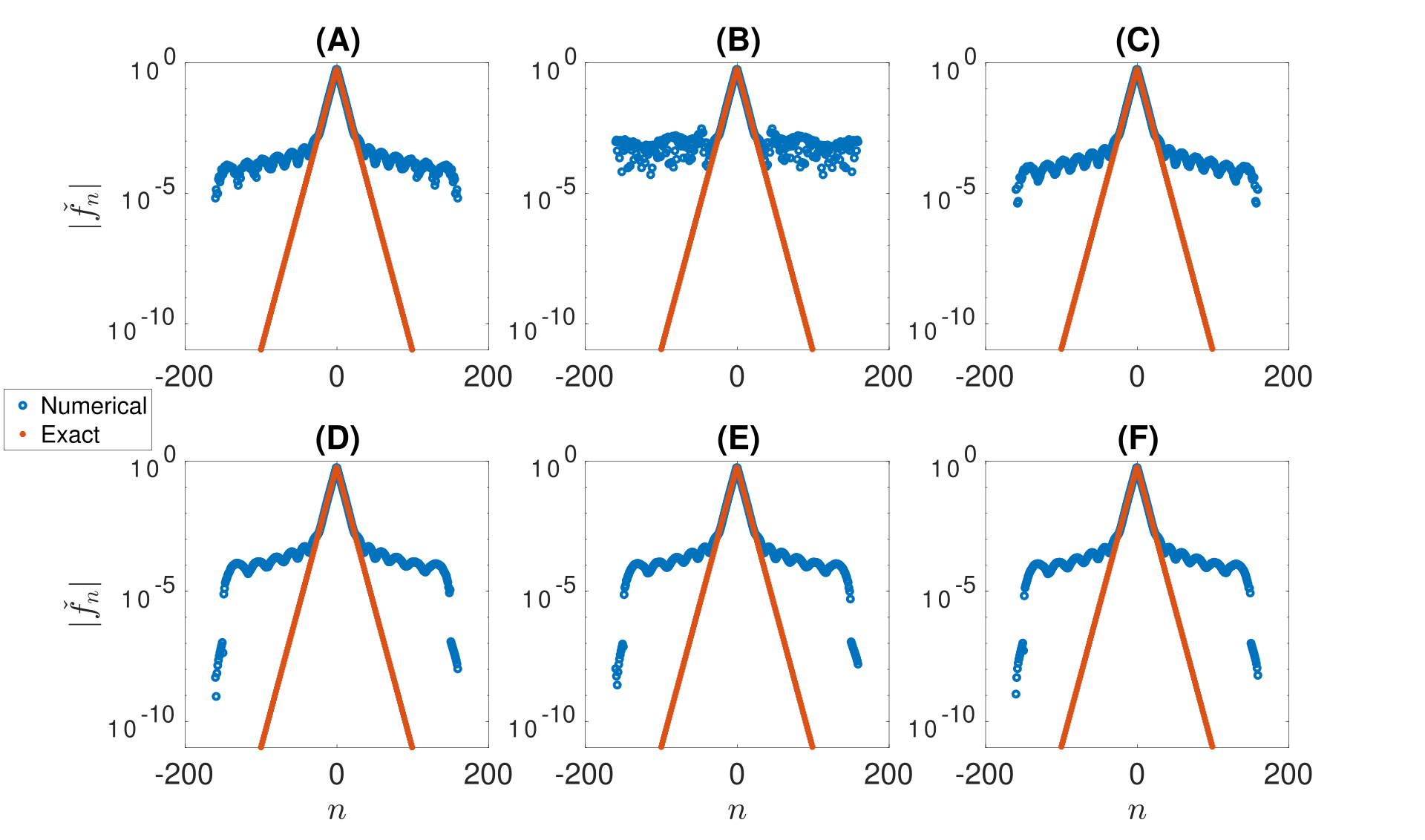}
   \caption{
   (A)-(C) 
   Modulus of the Ch-YSH numerical and exact Ch coefficients  {\it without padding}  for the exact solution given in (\ref{peregrine}) with $N = 320$ at final time 
   $t = -2+j\Delta t \approx 2$,  for (A) $\Delta  t=0.01(0.99)^{33}$, $j=557$, (B) $\Delta t =0.01(0.99)^{34}$, $j=563$, and (C) $\Delta t =0.01(0.99)^{35}$, $j=569$. (D)-(F) Modulus of the  Ch-YSH numerical and exact Ch coefficients  {\it with padding} for (D) $\Delta  t=0.01(0.99)^{33}$, $j=557$, (E) $\Delta t =0.01(0.99)^{34}$, $j=563$, and (F) $\Delta t =0.01(0.99)^{35}$, $j=569$; all other parameters are the same as the top row.   \label{nls_alias_end}}
   \end{figure}

To understand the source of this unexpectedly large error, we examine the evolution of  Ch coefficients for 
the function $u(x,t) - 1$. In Fig.~\ref{nls_alias_end} we display 
the Ch coefficients 
at $t=2$ (final time) for different choices of $\Delta t$. 
In Fig.~\ref{nls_alias_end}(B) the Ch coefficients for the problematic case are observed to be unexpectedly large at large modal values, i.e. at large values of $|n|$. 
The Ch functions $\phi_n(x)$ at large values of $|n|$ correspond to highly oscillatory functions (see Fig.~\ref{plot_Ch_fcns}).
 In contrast, the Ch coefficient profiles in Figs.~\ref{nls_alias_end}(A)  and \ref{nls_alias_end}(C)  appear to remain relatively small at large values of $|n|$.

The discrepancy in the  Ch coefficients for $\Delta t= 
0.01(0.99)^{34}$ compared to the others appears to be an aliasing-type error. For certain values of $\Delta t$, high Ch modes (large $n$) are unexpectedly large. This is a result of the nonlinearity in the NLS equation. 
For Fourier spectral methods, there are a myriad of techniques for de-aliasing, see \cite{Canuto2006}. 
One of the simplest techniques 
is that of padding. Spectral padding  
involves setting a certain number of high Ch modes (large $|n|$) 
to zero every time-step. 
Explicitly, we apply the following filter once at each linear solve in the Ch-YSH method
\begin{equation}
\label{padding_algorithm}
\tilde{u}_n = 
 \begin{cases} 
      \Check{u}_n , &  |n| < \frac{N}{2}-P \\
      0 , &  \text{otherwise}  \\
   \end{cases} ,
\end{equation}
for $P \ge 0$ whereby all Ch modes with corresponding modenumber $|n|$ greater than or equal to $N/2 - P$ ($N$ is even) are effectively truncated.

   \begin{figure}
    \centering
   \includegraphics[scale=0.3]{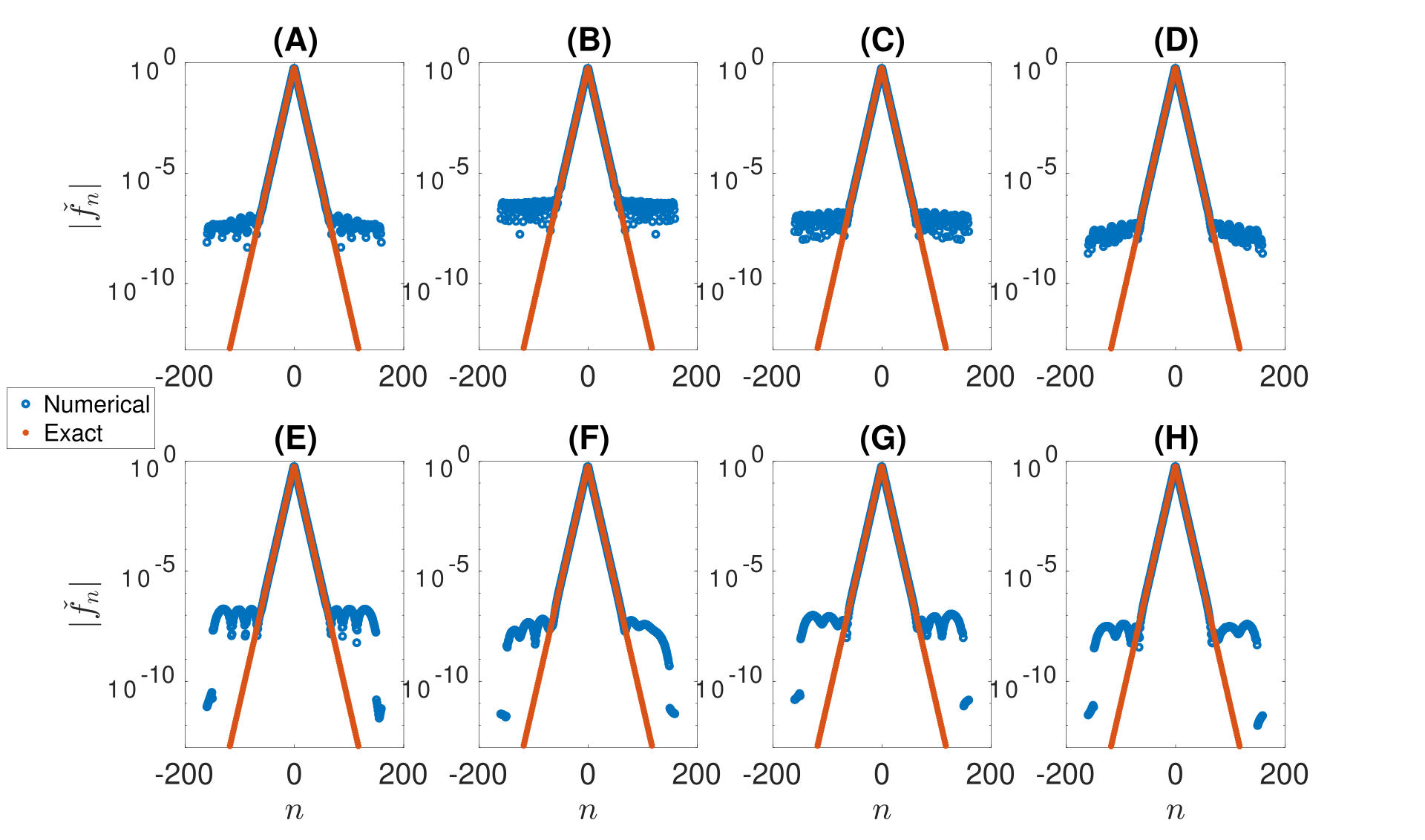}
   \caption{
(A)-(D) Modulus of the Ch-RKN numerical and exact Ch coefficients  {\it without padding}  for the exact solution given in (\ref{peregrine}) with $N = 320$ Ch modes at final time 
$t = -2+j\Delta t \approx 2$,  for (A) $\Delta  t=0.01(0.99)^{24}$, $j=509$, (B) $\Delta t =0.01(0.99)^{27}$, $j=525$, (C) $\Delta t =0.01(0.99)^{30}$, $j=541$, and (D) $\Delta t =0.01(0.99)^{33}$, $j=557$. (D)-(H) Modulus of the Ch-RKN numerical and exact Ch coefficients  {\it with padding}  for (E) $\Delta  t=0.01(0.99)^{24}$, $j=509$, (F) $\Delta t =0.01(0.99)^{27}$, $j=525$, (G) $\Delta t =0.01(0.99)^{30}$, $j=541$, and (H) $\Delta t =0.01(0.99)^{33}$, $j=557$; all other parameters are the same as the top row.
  \label{rknnls_alias_end}}
   \end{figure}


The convergence rate for the padded Ch-YSH method is shown  in Fig.~\ref{nls_alias}(B). 
Here we take a pad of $P = 10$ which zeros out 21 high frequency 
modes. This approach is found to prevent the growing tails observed in Fig.~\ref{nls_alias_end}(B) and preserve the fourth-order convergence rate (see bottom row of Fig.~\ref{nls_alias_end}). 
Unlike the unpadded case, we do not observe any sharp jumps in error for the $\Delta t$ values considered before.

The convergence curve for the RKN method in Fig.~\ref{nls_alias}(left) exhibits  local oscillations on a line that otherwise converges at the expected rate.  
An example of the oscillation occurs  at  $\jtc{\square}$: $\Delta t=0.01(0.99)^{24}$, $\jtc{\triangle}$: $\Delta t=0.01(0.99)^{27}$, $\jtc{\triangledown}$: $\Delta t=0.01(0.99)^{30}$ and $\jtc{\diamond}$: $\Delta t=0.01(0.99)^{33}$. Spectral padding can smooth out these oscillations without sacrificing the overall convergence rate 
of the method.

 In Fig.~\ref{rknnls_alias_end} we display 
the Ch coefficients of $u(x,t)-1$ 
at $t=2$ (final time) for different choices of $\Delta t$.
We observe that 
near 
the oscillation peak in Fig.~\ref{nls_alias} the corresponding Ch coefficients in Fig.~\ref{rknnls_alias_end}(B),(C) have noticeably larger values 
at large modal values ($|n| \gg 1$), as opposed to the valley of the oscillation Fig.~\ref{rknnls_alias_end}(A),(D). The convergence rate for the padded Ch-RKN method is shown  in Fig.~\ref{nls_alias}(B). 
Here the pad is still $P = 10$ which zeros out 21 high frequency 
modes. This approach is found to prevent the 
tails observed in Fig.~\ref{rknnls_alias_end}(B),(C). The convergence rate is at least 
fourth-order.

\end{document}